\documentclass{ptephy_v1}%%%%%% to generate preprint number with ptep logo

\preprintnumber{XXXX-XXXX} %%% %%% Insert preprint number here

\usepackage{graphics}% for dealing with figures.
\usepackage{subfig}% for getting the subfigures e.g., "Figure 1a and 1b" etc.
\begin{document}

\title{Hyper-track selector nuclear emulsion readout system aimed at scanning an area of one thousand square meters}

%%%% To generate auto affiliation numbers please use \author{}\affil{} command

\author[1\thanks{Present Address: Physics Department, Faculty of Education, Gifu University, 1-1 Yanagido, Gifu 501--1193, Japan}*]{Masahiro Yoshimoto}
\author[]{Toshiyuki Nakano}
\author[]{Ryosuke Komatani}
\author[]{Hiroaki Kawahara}
\affil{Graduate School of Science, Nagoya University, Furo-cho, Chikusa-ku, Nagoya 464--8602, Japan \email{yoshimoto@flab.phys.nagoya-u.ac.jp}}

% \affil{Insert second author address here}

% \author{Insert third author name here}
% \author[3]{Insert fourth author name here} %%% Use optional bracket [3] to change the respective address
% \affil{Insert third author address here}

% \author{Insert last author name here\thanks{These authors contributed equally to this work}}
% \affil{Insert last author address here}

%%% To include the collaborator name... Please use the command "\collaborator"
%%% For example: \collaborator{ATLAS Collaboration}

% \linenumbers %削除すべき
\begin{abstract}%
Automatic nuclear emulsion readout systems have seen remarkable progress since the original idea was developed almost 40 years ago. 
After the success of its full application to a large-scale neutrino experiment, OPERA, a much faster readout system, the hyper-track selector (HTS), has been developed. 
HTS, which has an extremely wide-field objective lens, reached a scanning speed of 4700~cm$^2$/h, which is nearly 100 times faster than the previous system and therefore strongly promotes many new experimental projects. 
We will describe the concept, specifications, system structure, and achieved performance in this paper.
\end{abstract}

\subjectindex{H01, H34}

\maketitle

\section{Introduction}
%%%% Nuclear emulsion %%%%
Nuclear emulsion is a three-dimensional particle tracking detector with a submicron spatial resolution.
It has been used for cosmic ray research and has contributed to the discoveries of $\pi$-meson~\cite{Brown:1949mj} and charmed particles~\cite{niu1971possible}.
To recognize the recorded trajectory with submicron accuracy in those cosmic-ray experiments, the emulsion films were scanned by eye using microscopes.
Consequently, this time-consuming method strongly limited the statistics.
%%%% Automatic nuclear emulsion %%%%
To improve this situation, an automatic nuclear emulsion readout system, named Track Selector, has been developed in Nagoya University ~\cite{niwa1974auto}\cite{Aoki:1989uk}\cite{nakano1997automatic} and applied to many high-energy experiments, WA75~\cite{Albanese:1985wk}, E653~\cite{Kodama:1990zy}, E176~\cite{Aoki:1991ip}, CHORUS~\cite{Eskut:1997ar}, and DONUT~\cite{Kodama:2000mp}.
Especially in DONUT, the readout system has discovered the $\tau$ neutrino.
Thanks to the experiment, the validity of automatic nuclear emulsion readout systems was demonstrated.
%%%% Previous system S-UTS and OPERA experiment %%%%
In 2006, S-UTS with a readout speed of $\sim$10~m$^{2}$/year/system was released~\cite{Morishima:2010zz} to organize a large-scale neutrino oscillation experiment, OPERA~\cite{Acquafredda:2006ki}.
Adding to S-UTS, a similar readout system was also developed in Europe~\cite{Armenise:2005yh}.
In the OPERA analysis over more than five years, the scanning area for those readout systems reached the order of 100~m$^2$, and consequently, the systems have contributed to the discovery of $\nu_{\mu} \rightarrow \nu_{\tau}$ oscillation in the appearance mode~\cite{Agafonova:2014ptn}\cite{Agafonova:2015jxn}.

The success of OPERA experiment---i.e., large-area application of nuclear emulsion and automatic nuclear emulsion readout systems---has led to several applications of nuclear emulsion technology---e.g., GRAINE project, NINJA project, and cosmic-ray muon radiography project.
%%%% GRAINE %%%%
The GRAINE project~\cite{Aoki:2012nn}\cite{Takahashi:2016xsf} aims at cosmic gamma-ray observation with a balloon bone telescope.
The angular resolution of nuclear emulsion is one order greater than that of Fermi LAT~\cite{atwood2009large}; hence, the angular resolution allows the telescope to take sharper images of gamma-ray emission objects such as supernova remnants.
The scientific observation will be performed with multiple flights and a large aperture of $\sim$10~m$^2$ to be configured on a balloon.
Test flights were carried out in 2011 and 2015, and observational flights using total area of 1000~m$^2$ nuclear emulsion are within our scope.
%%%% NINJA %%%%
The NINJA project~\cite{Fukuda:2017clt} aims at the precise measurement of low-energy neutrino-nucleus interactions and sterile neutrino search at J-PARC.
Only nuclear emulsion can detect short tracks and large angle tracks---i.e., low-energy charged particles such as recoil protons and nuclear evaporation fragments emitted to close to the 4$\pi$ region. 
%%%% Cosmic-ray Muon radiography %%%%
The cosmic-ray muon radiography project aims at observing the inner structures of large-scale objects including volcanos, iron furnaces, and nuclear power plants and archeological buildings including pyramids.
Nuclear emulsion has advantages in its compactness, including no requirement of power supply and large angular acceptance.
Nuclear emulsion also features scalability, which realizes almost 100 times more area than other technologies such as a scintillation detector under a given cost. 

To work with those requirements for the experiments, we have been developing the next-generation nuclear emulsion readout system aiming at achieving a readout speed of $\sim$1000~m$^2$/year---i.e., 100 times faster than S-UTS.
In this paper, we will present the Hyper Track Selector (HTS) concept, basic configuration, evaluation of key devices, track detection performance, achieved readout speed, and future prospects to achieve a much faster system.

\section{Automatic nuclear emulsion readout system and the concept of HTS}
\label{sec:Concept}
\begin{figure}[htbp]
\centering
\includegraphics[width=0.6\textwidth]{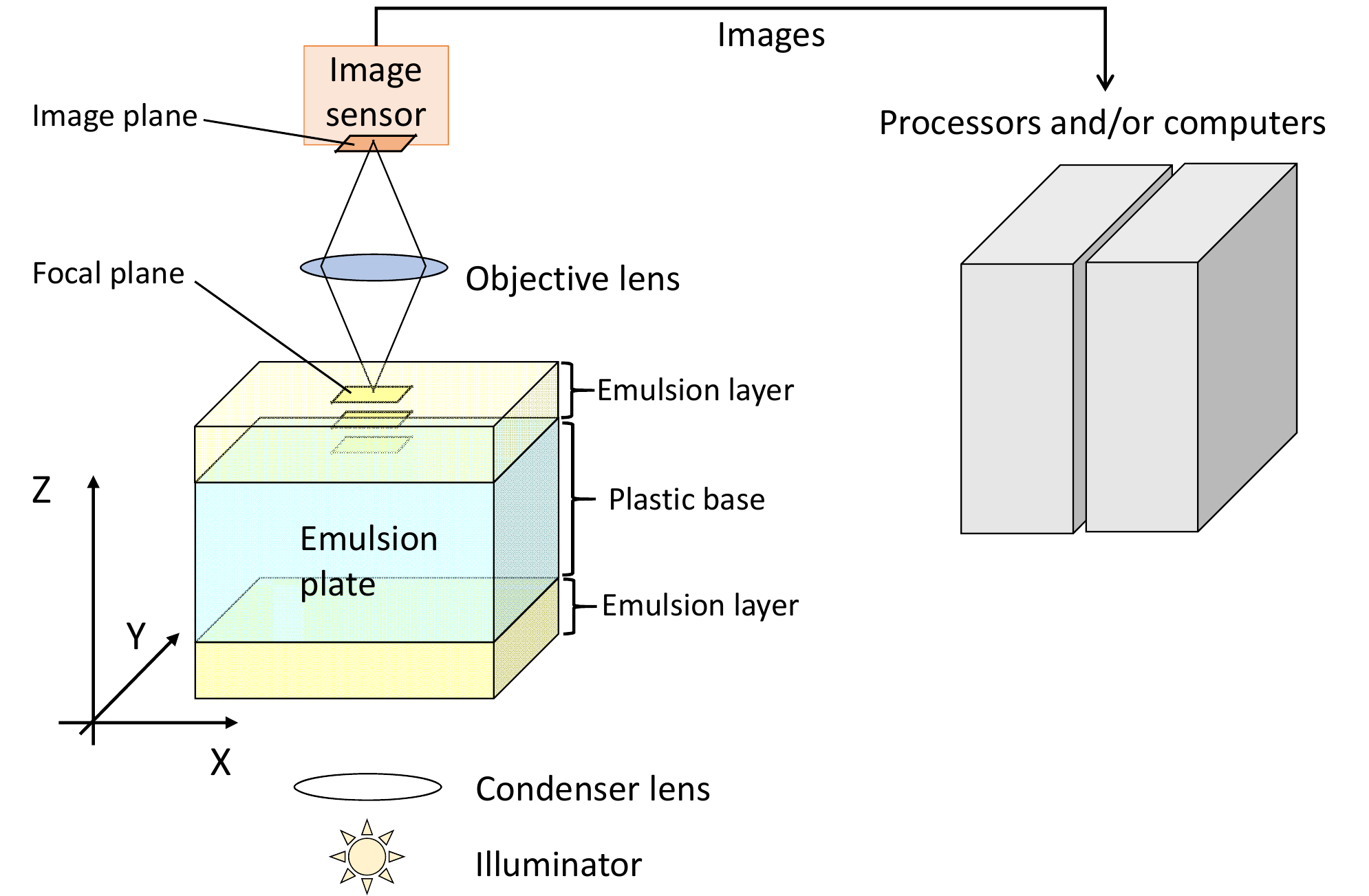}
\caption{Outline of a nuclear emulsion automatic readout system.}\label{fig:abst_system}
\end{figure}

In nuclear emulsion, trajectories of charged particles are recorded as three-dimensionally aligned dots of developed silver grains.
As shown in Fig.~\ref{fig:abst_system}, 
the readout system takes tomographic images in an emulsion layer and recognizes those linked grains across the tomographic images as a track.
A track-finding algorithm, ``Track Selector,'' especially searches straightly aligned dots---i.e., high-momentum tracks.

Normally, the system is composed of a three-axis stage to scan the emulsion film, an objective lens to magnify the nuclear emulsion images, an image sensor for capturing tomographic images, and dedicated processors and/or computers to process the acquired images and manage the total system. The objective lens and the image sensor should satisfy the requirement of submicron resolution. To increase the readout speed, it is necessary to improve the performance of each device in a well-balanced way.

\begin{figure}[htbp]
\centering
\includegraphics[width=0.6\textwidth]{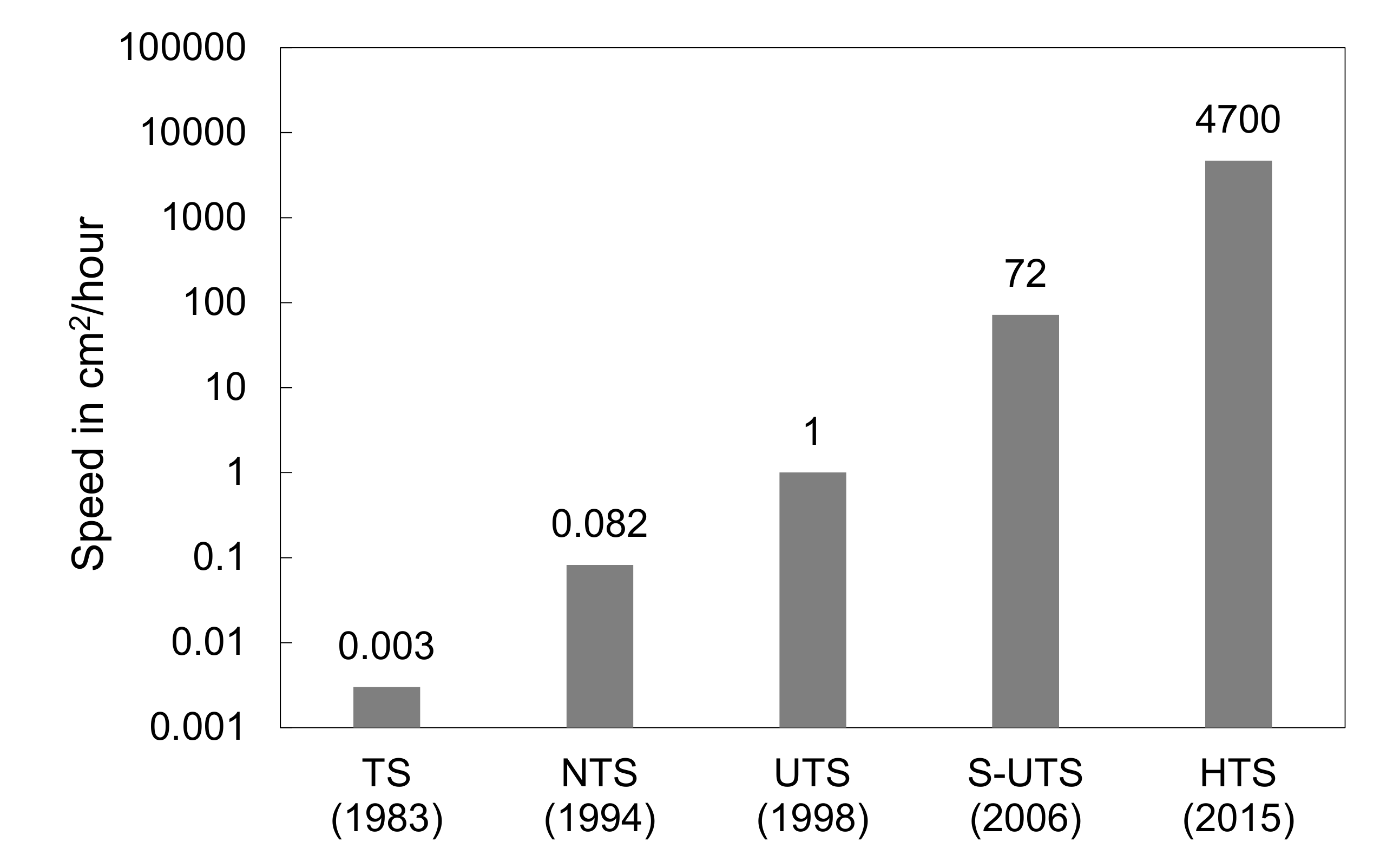}
\caption{Development of the scanning speed of the Track Selector series. 
The values of TS and NTS are the speeds at the same angular acceptance as UTS.}\label{fig:evolution}
\end{figure}

Fig.~\ref{fig:evolution} shows the development of the readout speed for the Track Selector series. 
The most appropriate digital technologies are used at each moment to increase the processing speed. 
Angle acceptance is also increased in every Track Selector series.
The TS and NTS have limited angle acceptance and then search a specific track in a view.
The UTS began to search all recorded tracks with the angular range of $\tan\theta\,<0.6$, where $\theta$ is the incident angle to a surface of an emulsion film.

The stage driving method has also been improved.
Before UTS, after the step movement of the $XY$-axis stage with motors to a targeted position, tomographic images were taken by changing the focal plane along the optical axis---i.e., $Z$-axis.
There is a dead time to wait for the vibration in the $XY$-axis stage to settle.
The vibration is caused by the acceleration and deceleration of the $XY$-axis stage; hence, so continuous stage movement has been adapted in S-UTS.
A dual piezo-driven objective lens has been installed; the $Z$-axis is used to change the focal plane, and the $X$-axis is used to cancel the $X$-axis stage movement.
As a result, the step and go movement of the $X$-axis stage has been removed.
The continuous stage movement overcame a bottleneck for the speedup.
In addition, a high-framerate camera of 3000~fps increased the readout frequency to 50~view/s, which is more than ten times faster than the previous system.
FPGA-based processors for high-speed image processing have also been developed to handle this high data rate.

Continuous-stage movement with higher-frequency movement of the objective lens seems unrealistic to achieve the further speedup, because the acceleration becomes $\sim$100~m/s$^2$ if we intend to achieve a readout frequency of 500~view/s.
Therefore, we changed the strategy from increasing the readout frequency to widening the field of view (FOV) for a new Track Selector, HTS.
To serve this purpose, we have developed a dedicated objective lens with an FOV of 5.1$\times$5.1~mm$^2$, which is approximately 500~times wider than the previous system.
To cover this wide FOV with submicron resolution, a mosaic camera system composed of 72 two-megapixel sensors has been developed.
The wide FOV is read by 72 sensors in parallel, and each captured image must be processed within the time comparable to that for stage movement.
Because the step motion was adapted again, the vibration bottleneck relapsed as described above.
Hence, a counter stage was introduced to cancel the movement and then hold the center of gravity for the stage, which seems to be the origin of the stage vibration.
A linear motor has been adapted to move the stage at a high acceleration and deceleration.
We also adapted consumer GPUs for the image processing and track recognition.
The GPUs are readily available, are widely used, and have design flexibility.

\begin{figure}[htbp]
\centering
\includegraphics[width=0.6\textwidth]{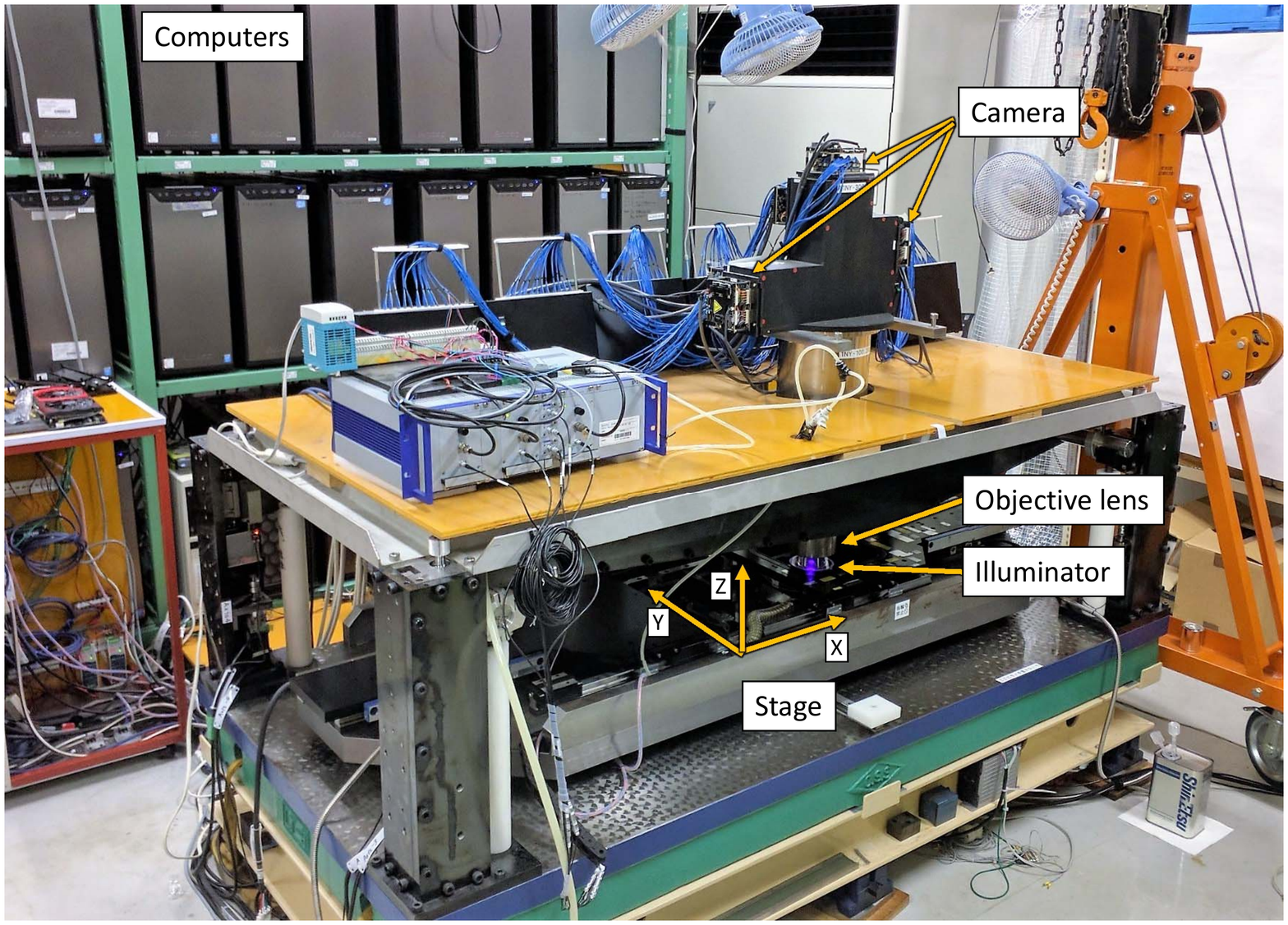}
\caption{Picture of HTS.}\label{fig:hts_full}
\end{figure}

Fig.~\ref{fig:hts_full} shows a picture of the HTS system, dedicated lens, camera, $XYZ$-axis stage mounted on a base, and computer cluster for image processing.
The details of each part will be described in the next section.

\section{The HTS system}
\subsection{Optics: Objective lens and illuminator}
\label{sec:optics}
\begin{figure}[htbp]
\centering
\subfloat[Objective lens and beam splitter]{\includegraphics[page=1,width=0.3\textwidth]{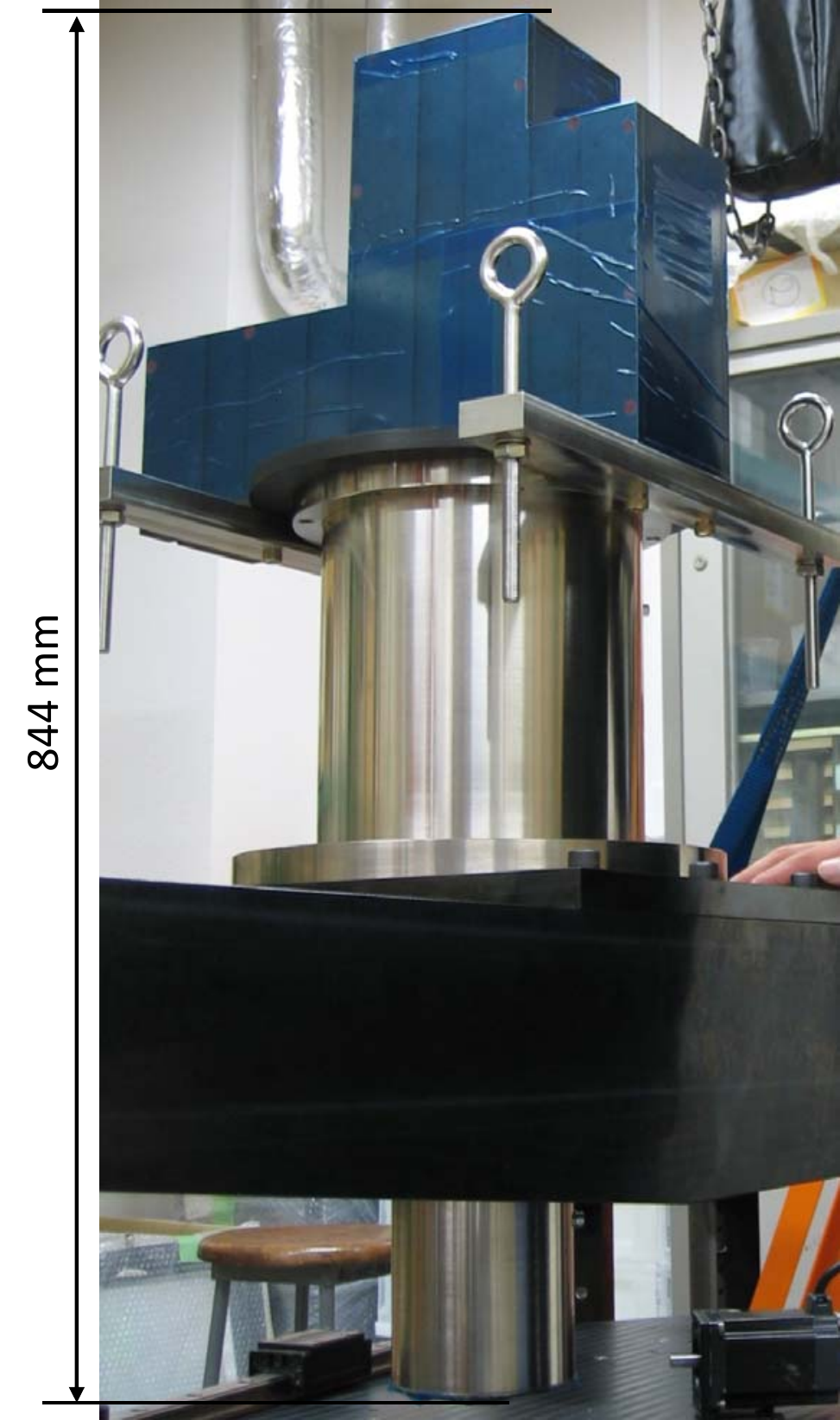}}
\hspace{1em}
\subfloat[Condenser lens for the illuminator]{\includegraphics[page=2,width=0.3\textwidth]{./figs_lens.pdf}}
\caption{Photos of the optics.}\label{fig:lens}
\end{figure}

The optics---i.e., objective lens, illuminator, and beam splitter---were designed and produced by KONICA MINOLTA Inc.
The specifications are shown in Table~\ref{tab:lens}, and the outlook is shown in Fig.~\ref{fig:lens}.
The FOV is 5.1$\times$5.1~mm$^2$, the magnification is 12.1$\times$, and the working distance is 1.5~mm.
Optical immersion oil with a refractive index of 1.505 must be inserted between the objective lens and the emulsion plate for optical matching.

To produce the wide view objective lens at a reasonable cost, the numerical aperture (NA) was limited to 0.65.
The NA was less than 0.85 of the previous objective lens with an FOV of 0.2$\times$0.2~mm$^2$.
To recover the degraded optical resolution by the smaller NA, light with wavelength $\lambda$ of 436~nm was utilized, which was shorter than the previous value of 550~nm.
No wavelength shorter than 400~nm can be used owing to the smaller transmittance of the material used for nuclear emulsion.
The lateral resolution defined by the Rayleigh criterion ($\delta x = 0.61\frac{\lambda}{\rm{NA}}$) became 410~nm.
The resolution was almost equal to that of the previous lens of 390~nm.
As a light source, a mercury xenon lamp was used to obtain enough brightness in the whole FOV, and bandpass optical filters centered at 436~nm with an FWHM of 10~nm were used to ignore chromatic aberration.
In addition, because of the smaller NA, the depth of field (DOF) 
($\delta z = 2\frac{\lambda}{\rm{NA^2}}$)---i.e., the resolution along the optical axis---became 2.1~$\rm{\mu m}$, 
which was approximately 1.5 times wider than that of the previous lens.

\begin{table}[htbp]
\centering
\begin{tabular}{|l|c|c|}
\hline
 & S-UTS & HTS\\
\hline
Objective lens &  & \\
\hline
Manufacture & TIYODA & KONICA MINOLTA\\
Magnification & 35$\times$ & 12.1$\times$ \\
Numerical aperture & 0.85 & 0.65 \\
Optimum wavelength & Green (550~nm) & Blue (436~nm) \\
Working distance & 1.1~mm & 1.5~mm \\
Depth of field & 1.5~$\rm{\mu m}$ & 2.1~$\rm{\mu m}$ \\
Field of view & 0.230$\times$0.228~mm$^2$ & 5.1$\times$5.1~mm$^2$\\
\hline
Illuminator & & \\
\hline
Condenser NA & 0.85 & 0.66 \\ 
Light source & Hg-Xe lamp & Hg-Xe lamp\\
Filter & Green filter & 436$ \pm $10 nm \\
\hline
\end{tabular}
\caption{Specification comparison of the objective lens and illuminator between S-UTS~\cite{Morishima:2010zz} and HTS. The depth of field is defined as $\delta z = 2\frac{\lambda}{\rm{NA^2}}$.}
\label{tab:lens}
\end{table}

\subsection{Camera}
HTS uses 72 image sensors. 
As shown in Table~\ref{tab:image_sensor}, one sensor has 2.2~megapixels (H2048$\times$V1088 pixels).
The physical pixel pitch of 5.5~$\rm{\mu m}$ corresponds to 0.45~$\rm{\mu m}$ on the object.
This value of 0.45~$\rm{\mu m}$ is approximately the diffraction limit of 0.41~$\rm{\mu m}$ as described in section \ref{sec:optics}.

\begin{figure}[htbp]
\centering
\includegraphics[width=0.5\textwidth]{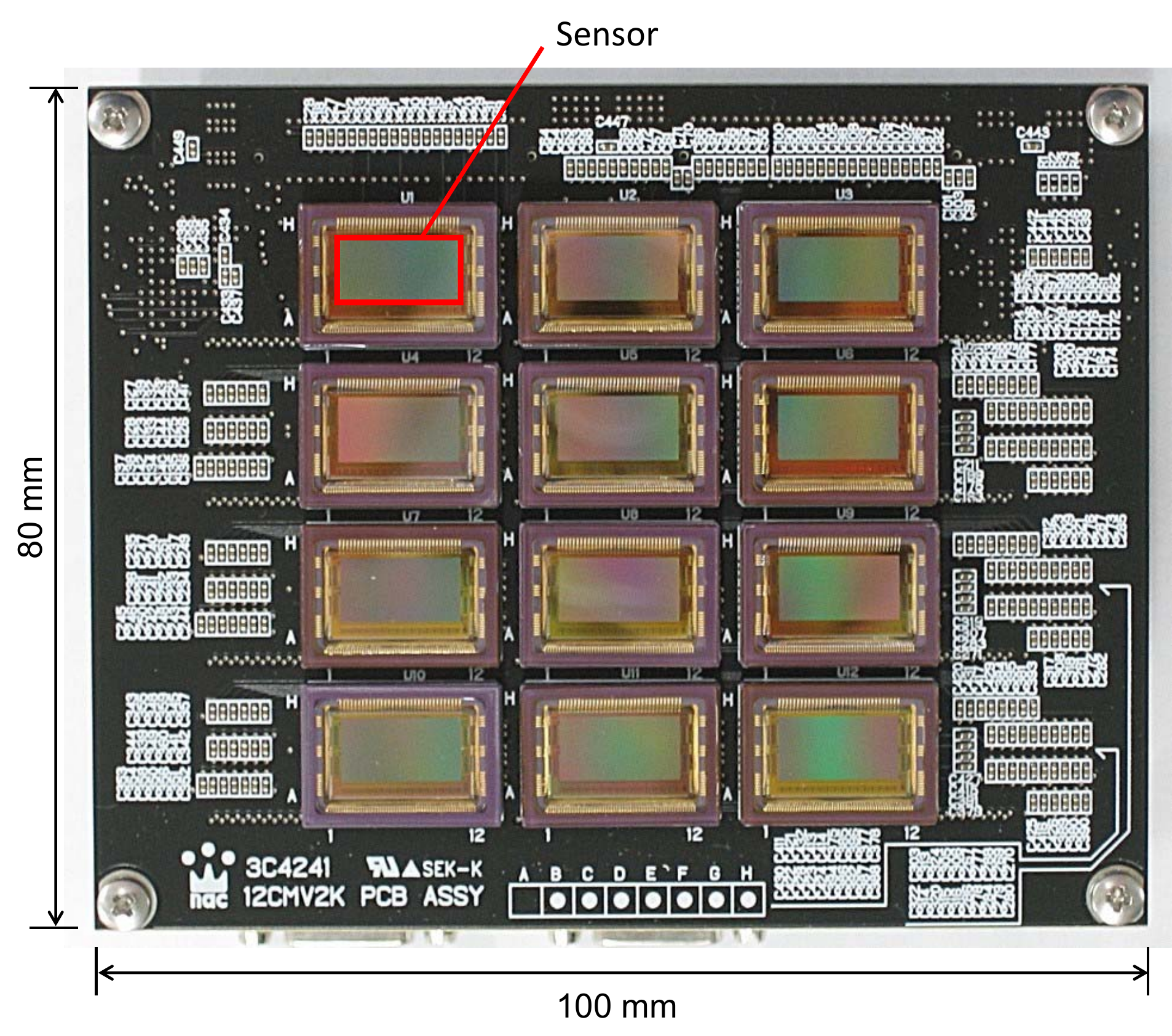}
\caption{Photo of mosaic camera unit. Twelve image sensors are mounted on one unit. The red bold frame shows the sensitive part of a sensor.}
\label{fig:camera}
\end{figure}

\begin{figure}[htbp]
\centering
\includegraphics[width=0.6\textwidth]{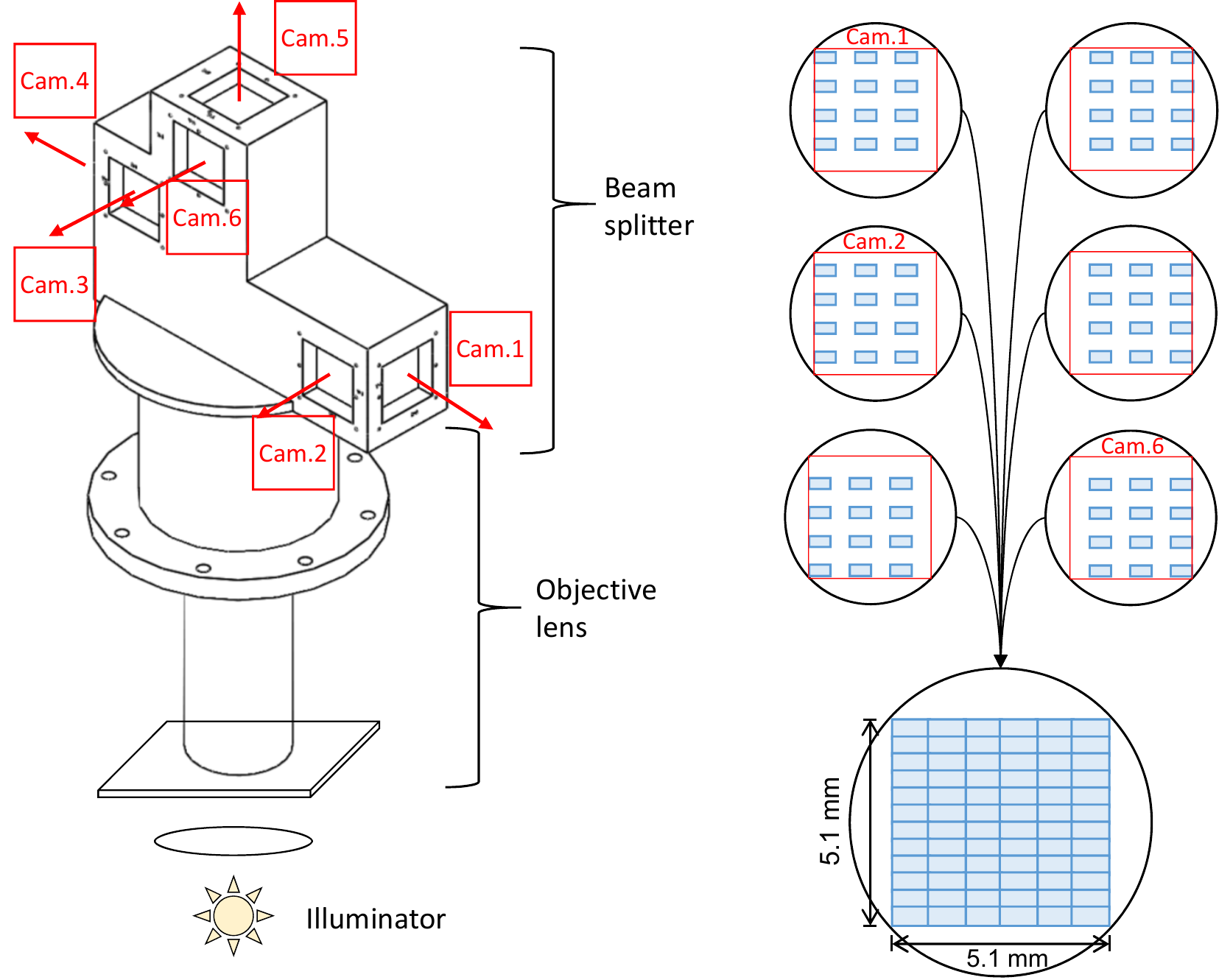}
\caption{Schematic view of the beam splitter unit with six windows (left) and the FOV configuration constructed with six mosaic camera units (right).}
\label{fig:lens_camera}
\end{figure}

Six mosaic camera modules, in which 12 image sensors are arranged respectively as shown in Fig.~\ref{fig:camera},
are installed on the six image planes divided by the beam splitter.
The sensors that compose the camera cover different parts of the 5.1$\times$5.1~mm$^2$ FOV as shown in Fig.~\ref{fig:lens_camera} 
to complement the dead spaces of the other cameras.
Each sensor has overlap greater than 50~$\rm{\mu m}$ (at the object) with the adjacent sensor 
to achieve tolerance for the alignment error of the sensor and camera and avoid missing the trajectory at the sensor edge.

The framerate of the sensor is 300~fps which is ten times slower than that of S-UTS.
The total image transfer rate is 48~GB/s, which is 60 times larger than that of S-UTS.

\begin{table}[htbp]
\centering
\begin{tabular}{|l|c|c|}
\hline
Image sensor  & S-UTS & HTS \\
\hline
Product name & MEMCAMfx RX-6 custom & CMV2000\\
Image sensor & CMOS & CMOS\\
Resolution & 512~$\times$~508~pixels & 2048~$\times$~1088~pixels \\
Framerate & 3000~fps & 300~fps \\
Pixel pitch & 16~$\rm{\mu m}$~$\times$~16~$\rm{\mu m}$ & 5.5~$\rm{\mu m}$~$\times$~5.5~$\rm{\mu m}$ \\
Electronic shutter & Available & Available \\
\hline
\end{tabular}
\caption{Specification comparison of the image sensor between S-UTS~\cite{Morishima:2010zz} and HTS.}\label{tab:image_sensor}
\end{table}

\subsection{$XYZ$-axis stage}

\subsubsection{$XY$-axis stages}
\begin{figure}[htbp]
\centering
\includegraphics[width=0.8\textwidth]{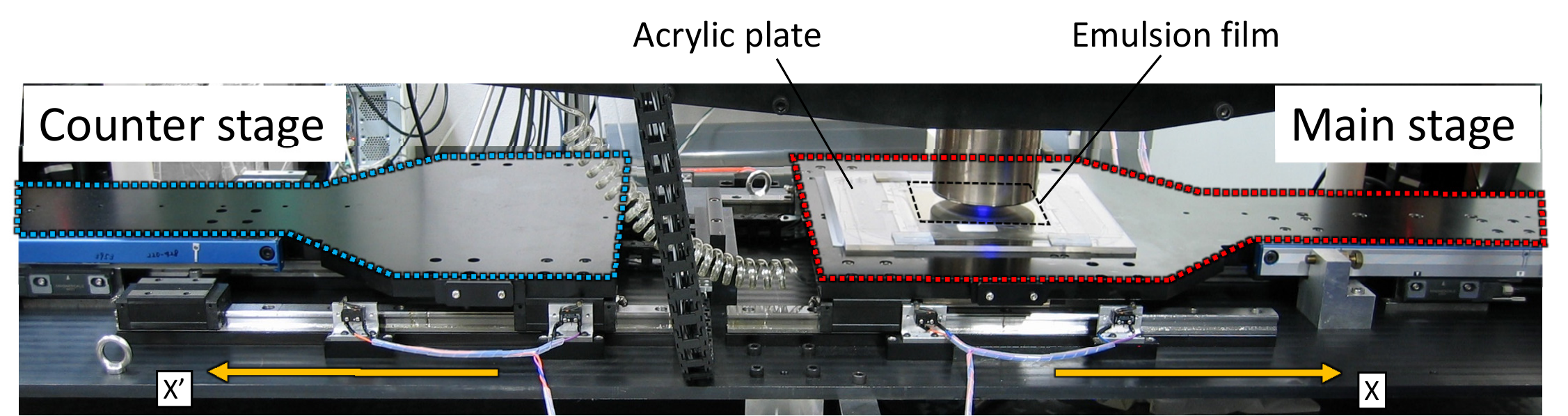}
\caption{Configuration of $X$-axis stage. 
The main stage and the counter stage are installed in line and are driven in the opposite direction to maintain the center of gravity at rest.}
\label{fig:main_counter_stage}
\end{figure}

The $XY$ stage is set on a metal surface table whose size is 2$\times$1~m$^2$ and weight is approximately 700~kg.
The stage stroke is matched up to the OPERA film whose size is 125$\times$100~mm$^2$~\cite{Nakamura:2006xs}. 
The $X$ direction is designed to be the main scanning direction, and the stroke is set to be 130~mm.
The $X$ stage is driven by a linear motor that can drive 5-mm step movements of the 35-kg $X$ stage in 20~ms.
As shown in Fig.~\ref{fig:main_counter_stage}, a counter stage is installed, and then the counter stage is driven to suppress the vibration caused by the movement of the system's center of gravity.
The stroke of the $Y$ stage is set to be 100~mm.
The $Y$ stage is driven by a rotary motor and a ball screw.
Three linear encoders are installed on each stage (two on $X$ and one on $Y$), and the encoders are used to monitor the stage position and feedback to each actuator.

\subsubsection{$Z$-axis stage}

The $Z$ stage has two driving mechanics---i.e., coarse lens movement and fine film movement.
The coarse lens movement drives the lens and its support, weighing 200~kg, to change the emulsion layers and exchange the scanning films.
The coarse movement consists of four rotary motors equipped with rotary encoders and ball screws, and the stroke is set to be 20~mm.
The fine film movement drives the film to acquire tomographic images of an emulsion layer.
The fine movement consists of piezo actuators with linear encoders for the feedback of the $Z$ position.

\subsection{Computer clusters}
\subsubsection{Computer configuration}
\begin{figure}[htbp]
\centering
\includegraphics[page=1,width=0.6\textwidth]{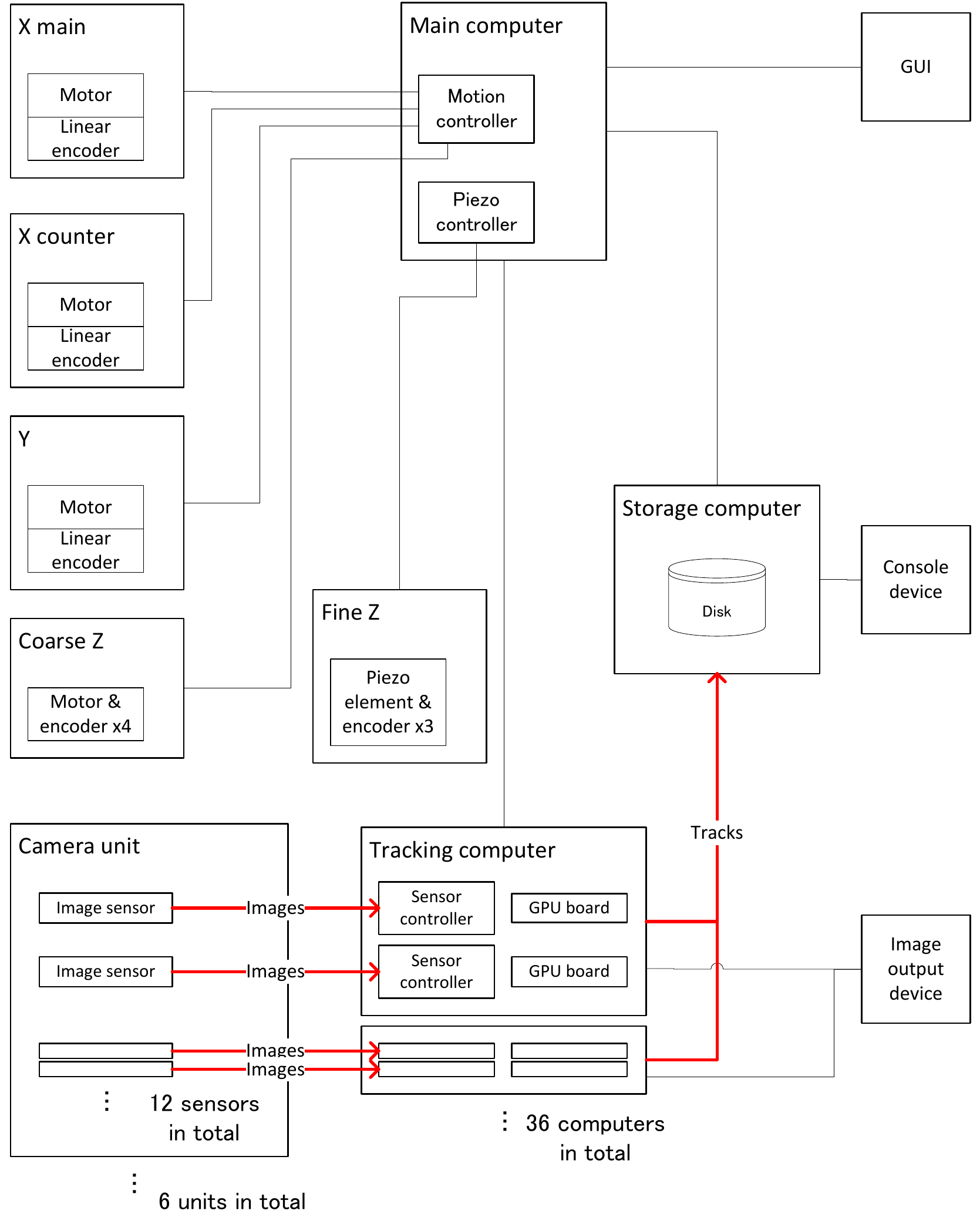}
\caption{Schematic view of the HTS system configuration.}\label{fig:hts_hardware}
\end{figure}

The HTS system configuration is shown in Fig.~\ref{fig:hts_hardware}.
The system is controlled by 38 computers: one main computer, one data storage computer, and 36 tracking computers.

The main computer controls the stage movement through a motion controller and piezo controller.
There are 72 image sensors in 6 camera units covering a 5.1$\times$5.1~mm$^2$ FOV.
Each sensor is connected to a sensor controller, and one GPU board (NVIDIA GeForce GTX680 or higher) is used for image processing and track recognition per sensor.
Two sensor controllers and two GPU boards are mounted on one tracking computer.
Thirty-six tracking computers then cover all 72 sensors.

\begin{figure}[htbp]
\centering
\includegraphics[page=3,width=0.6\textwidth]{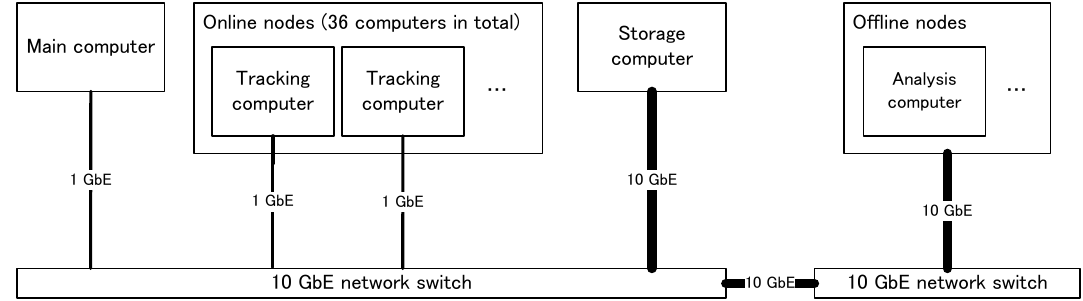}
\caption{Network configuration of the HTS system.}\label{fig:network_system}
\end{figure}

The network configuration of those computers is shown in Fig.~\ref{fig:network_system}.
The main computer, tracking computers, and data storage computer are connected through Ethernet.
The main computer and tracking computers are connected through Gigabit Ethernet (GbE).
The main computer triggers the data capture sequence of the tracking computers and receives feedback information including the number of recognized grains from the tracking computers to monitor the status of data capture.
Recognized track data are transferred from tracking computers to a storage computer.
The storage computer is connected through 10~GbE to simultaneously transfer the data from 36 tracking computers.
After storing the data on the storage computer, the analysis computers reconstruct tracks from the output data for offline analysis.
The analysis computer is also connected with 10~GbE to copy the data from the data storage computer.

\subsubsection{Software configuration}
\begin{figure}[htbp]
\centering
\includegraphics[page=2,width=0.6\textwidth]{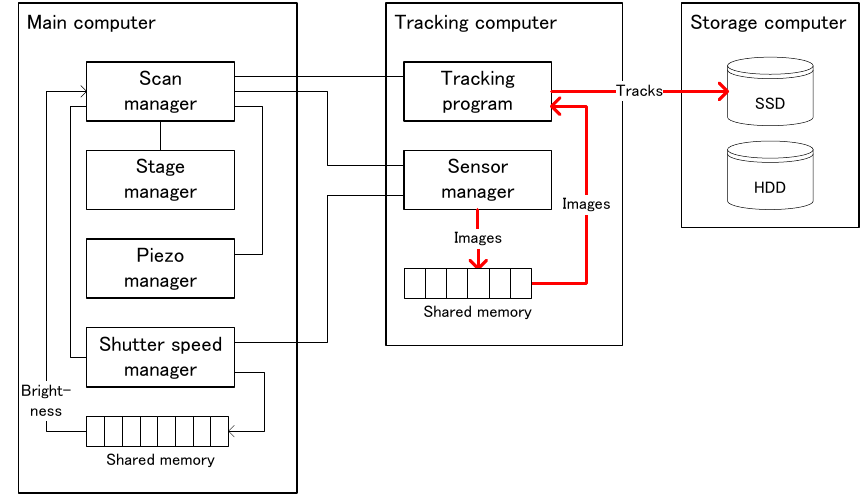}
\caption{Software configuration of the HTS system.}\label{fig:software_system}
\end{figure}

The software configuration is shown in Fig.~\ref{fig:software_system}.
A scan manager, stage manager, piezo manager, and shutter manager operate on the main computer.
The track recognition program and sensor manager operate on the tracking computers.
The images used in the tracking recognition program are transferred from the sensor manager via the shared memory.

\subsection{Track recognition procedure}

Twenty-two tomographic images are taken by changing the focal plane through the emulsion layer,
and sixteen successive images in the emulsion layer are used for track recognition.
A typical thickness of an emulsion layer is 60~$\rm{\mu m}$.

\subsubsection{Image processing}
Image processing applied to the original images has two steps.
The two steps are performed on the GPUs using the OpenCV library (version 2.4).

\begin{figure}[htbp]
\centering
\includegraphics[width=0.6\textwidth]{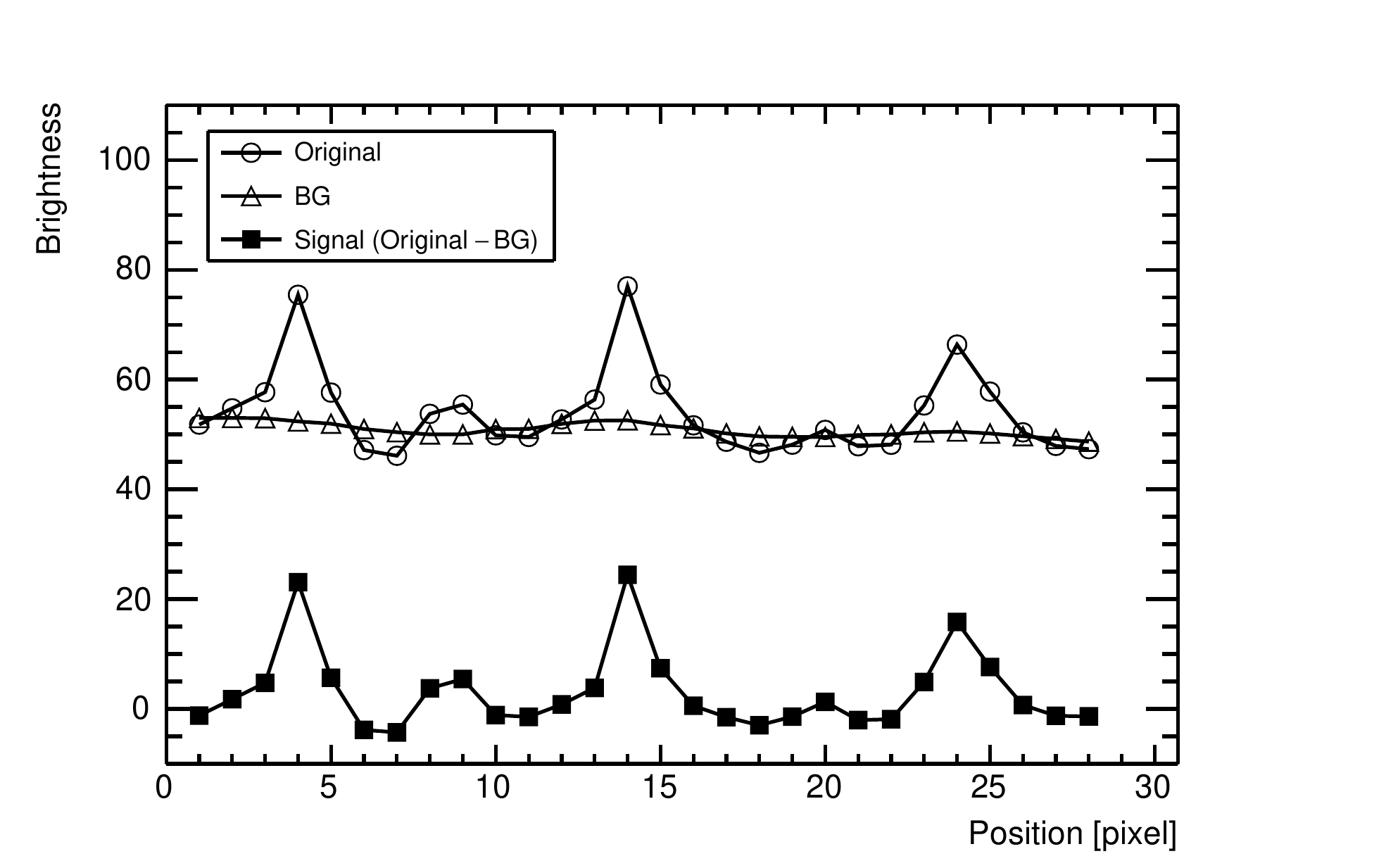}
\caption{
    Schematic view of the image processing to extract grains.
    The low-frequency background $BG$ is extracted from the original image.
    Grains appear as the peaks in the signal, which is the $BG$ subtracted from the original.}\label{fig:original_bg_signal}
\end{figure}

The first step is an extraction of focused grains on the images.
A low-pass filter is applied to the original image to create a blurred background image.
A grain has a typical size of 0.5~$\rm{\mu m}$, which corresponds to approximately one pixel on the image.
Thus, the cutoff frequency of the low-pass filter should be approximately 1/(10 pixels).
The background brightness $B_{BG}(x,y)$ for a pixel $(x,y)$ is calculated by averaging the weighted brightness of $N$$\times$$N$ pixels 
surrounding a specified pixel of original brightness $B_{Original}$ with a matrix---i.e., a kernel $K$.
\begin{equation*}
B_{BG}=\sum_{i=1}^{N} \sum_{j=1}^{N}B_{Original}\left(x+i,y+j\right)\cdot K(i,j).
\label{equ:convolution}
\end{equation*}
We used a cutoff frequency of 1/(12.5 pixels) and a kernel with 15$\times$15 matrices.
By subtracting this averaged image from the original image, an image with a high-frequency component is obtained as shown in Fig.~\ref{fig:original_bg_signal}.
Finally, by applying a threshold cut, we obtained the grains on the images.

\begin{figure}[htbp]
\centering
\subfloat[No expansion]{\includegraphics[page=1,width=0.2\textwidth]{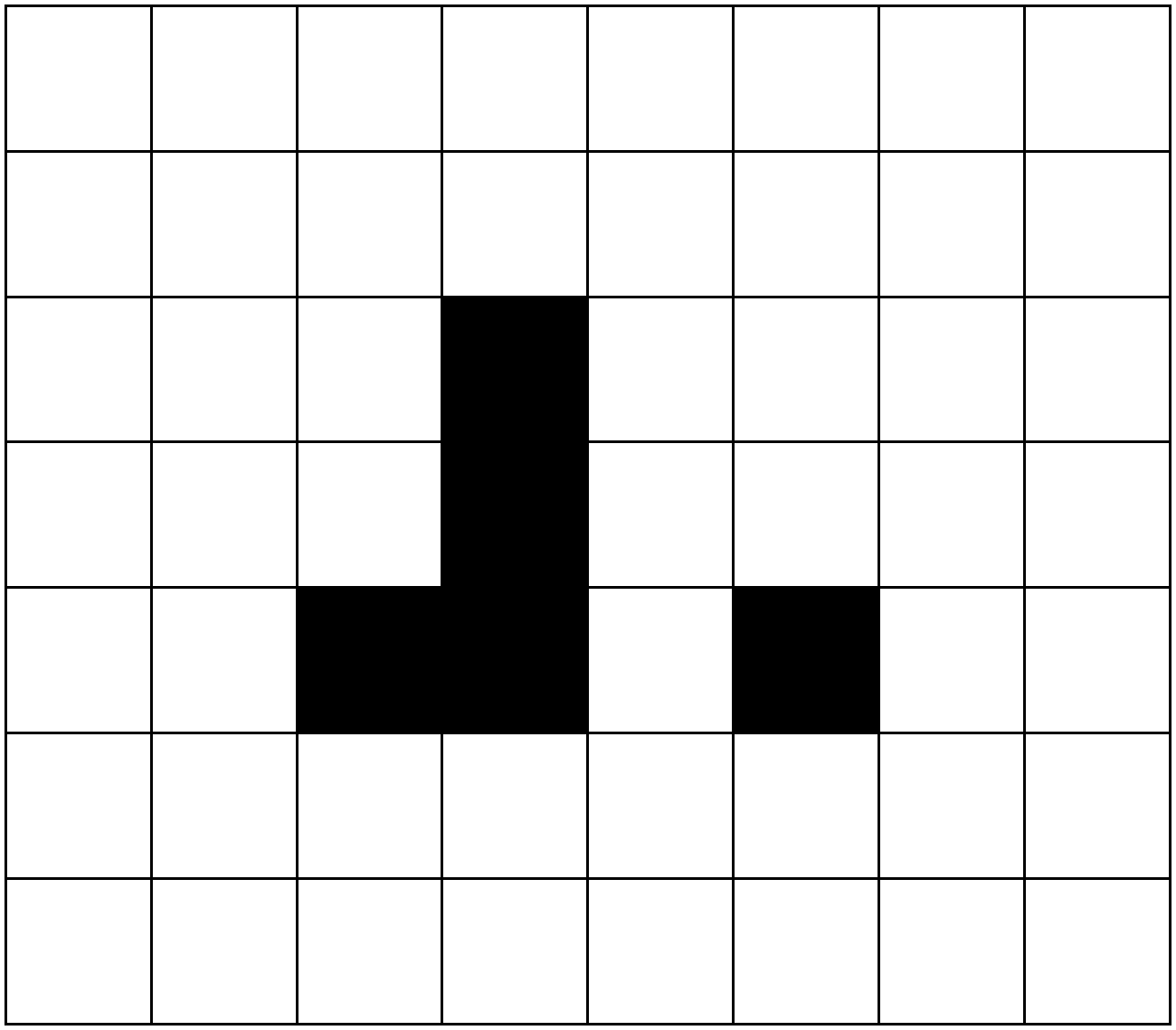}}
\hspace{1em}
\subfloat[2$\times$2 expansion]{\includegraphics[page=2,width=0.2\textwidth]{./figs_expansion.pdf}}
\hspace{1em}
\subfloat[3$\times$3 expansion]{\includegraphics[page=3,width=0.2\textwidth]{./figs_expansion.pdf}}
\caption{
    Schematic view of the image expansion.
    The original pixels and the expanded pixels are the black pixels and the gray pixels, respectively. }\label{fig:expansion}
\end{figure}

At the second step, elimination of off-focus grains spread over the images with different focal depth is applied.
This process was performed on the $BG$ image with the before-and-after images.
\begin{equation*}
B'_M=B_M-\frac{a}{2}\left(B_{M-1}+B_{M+1}\right).
\label{equ:zfilter}
\end{equation*}
The factor $a$ was adjusted so that the spread of focused grains was less than the interval of the images and $M$ is the number of tomographic images.
After those processes, a threshold cut is applied to obtain binary images.
Hit expansion of 2$\times$2 to up, right, and upper-right of the original hit pixels (Fig.~\ref{fig:expansion} (b)) 
% with a kernel {\footnotesize$\left(\begin{array}{ccc}0 & 0 & 0 \\0 & 1 & 1 \\0 & 1 & 1\end{array}\right)$} 
or 3$\times$3 to surround the original hit pixels (Fig.~\ref{fig:expansion} (c)) 
% with a kernel {\footnotesize$\left(\begin{array}{ccc}1 & 1 & 1 \\1 & 1 & 1 \\1 & 1 & 1\end{array}\right)$} 
is then carried out, because the size of the grain hit is normally smaller than the allowance of the trajectory recognition algorithm.
We usually utilized 2$\times$2 expansion to improve signal-to-noise ratio.

\subsubsection{Judgment on the success of the image acquisition}\label{sec:surface_recognition}
It should be determined whether the acquired tomographic images adequately cover an emulsion layer before moving to the next FOV.
In the standard setting, 22~images are captured; sixteen successive images are used for track recognition, and then six images are used as margin images. 

The number of grains (NOG) recognized in each image is calculated by image processing step 1.
The existence of the emulsion layer in the images is judged when the NOG is greater than a defined threshold.

\begin{figure}[htbp]
\centering
\includegraphics[width=0.6\textwidth]{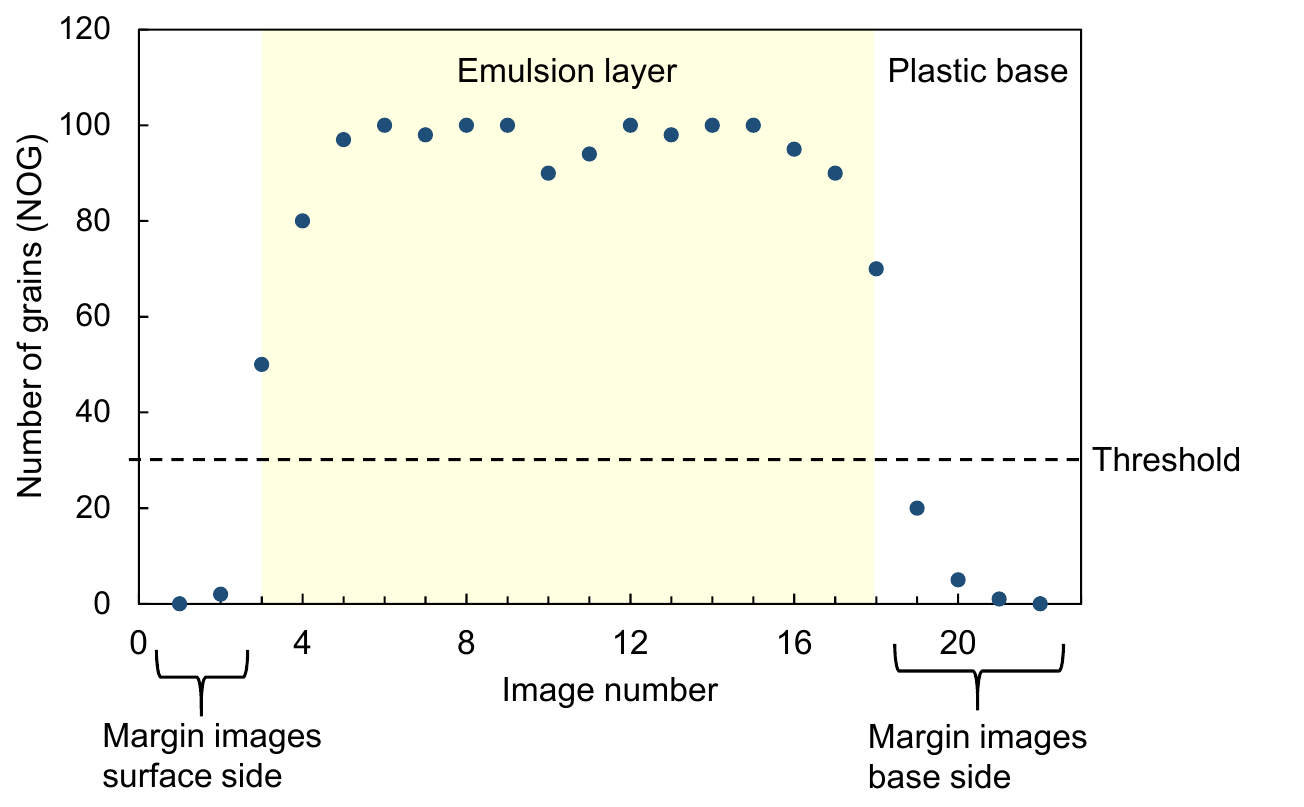}
\caption{Method to locate emulsion layer position among tomographic images.}\label{fig:judge_nog}
\end{figure}

The $Z$ positions of each sensor on the object are spreading in approximately two images at present even if the emulsion film surface is completely flat.
As shown in Fig.~\ref{fig:judge_nog}, using NOG information, the numbers of margin images on the base side and surface side are calculated for each sensor.
The two average values of the margin images over all sensors are calculated.
If either average value is less than 1, the image data acquisition is attempted again.
Moreover, if either average value is greater than 1 and less than or equal to 1.5, the $Z$ position to start image capture is corrected in the next FOV.
These treatments minimize the time loss to data recapture.

\subsubsection{Track recognition}
Straightly aligned grains through the sixteen binarized tomographic images are searched with the method described in the previous paper~\cite{Morishima:2010zz}.
The method, the so-called Track Selector algorithm, uses simple shift and sum functions.
The binary images are shifted to $X$ and $Y$ as the specified angle tracks become perpendicular to the focal plane and summed perpendicularly.
The summed value, called pulse height (PH), is an indicator of the track likelihood.
By setting the threshold, tracks with the specified angle are then recognized.
The process is repeated by changing the angle specification---i.e., scanning the required angular range.
The relative shift value from the first image to sixteenth image is given to be every three pixels ranging within $\pm$180~pixels for $X$ and $Y$ individually in the case of the standard condition.
The corresponding angle range is $\pm$53$^{\circ}$ in the case of a 60-$\rm{\mu m}$-thick emulsion layer.
Although the readout speed becomes slower, we can choose the relative shift pixel ranging within $\pm$360~pixels---i.e., the corresponding angle range of $\pm$70$^{\circ}$.
This tracking program can handle up to $\sim$10$^6$~tracks/cm$^2$ without extra processing time.
The speedup of the tracking programs installed in the GPU is one of the keys to realize HTS.
The details will be described in a future paper.

After applying threshold cut, clustering of the recognized tracks in angle and position space is applied because one track has multiple hits spreading over the angle and position space.
The clustering process is performed by the CPU of the tracking computer.
At this clustering process, the volume of the hits over the position space ($PHvol$) is also calculated, which represents the darkness of the track---i.e., the size of ionization energy loss per unit length ($dE/dx$).
The finally identified track is called a micro track and is the base for further track reconstruction.

\subsubsection{Diagram of the procedure}
\begin{figure}[htbp]
\centering
\includegraphics[page=1,width=0.6\textwidth]{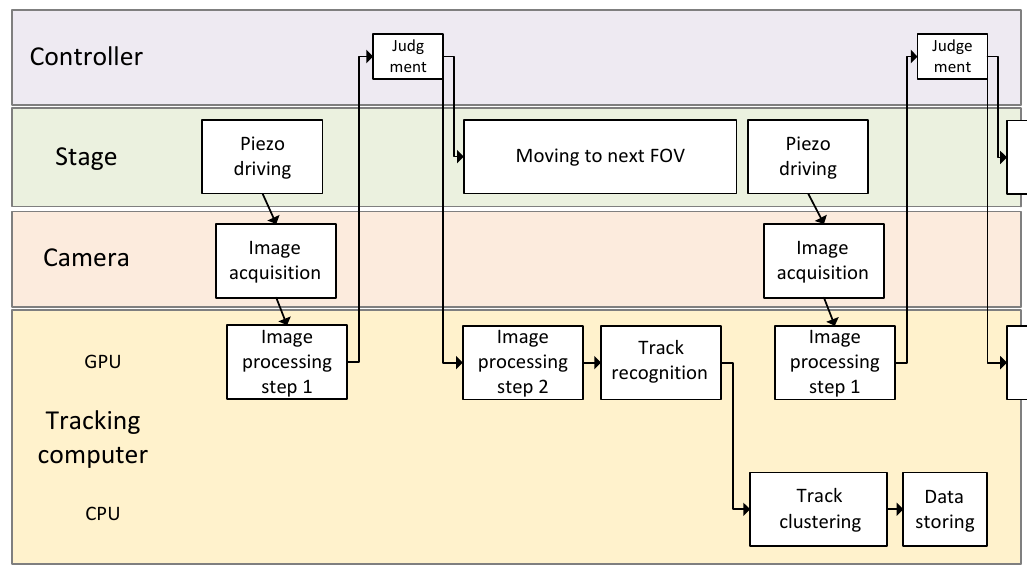}
\caption{Time sequence of each device activity.
    The length of the squares is not proportional to the actual time.}\label{fig:timechart}
\end{figure}

The time chart for scanning is shown in Fig.~\ref{fig:timechart}.
Except at the time of judgment, the stage, the camera, and the tracking computers work in parallel.
In the standard setting of track recognition,
the time for moving to the next view is longer than the total of image processing step 2 and track recognition.

\section{Sensor-to-sensor alignments}
Our unification of the 72 independent image sensors into one coordinate is the first such attempt in the history of nuclear emulsion readout systems.

The objective lens has a distortion such as ``radial distortion'' in its wide view of 5.1$\times$5.1~mm$^2$.
Although this distortion is thought to affect the position and angle of the recognized tracks, the effects can be ignored owing to the fine segmentations of the FOV into 72 sensors; 
an affine transformation parameter of each sensor can correct the magnification difference and position offset for each sensor.

The coordinate system of each sensor should be unified into the stage coordinate based on the encoders of the $XYZ$-axis stage.
The coordinate transformation from the sensor coordinate $(x, y)$ to the stage coordinate $(X, Y)$ is defined as 
\begin{equation*}
\left(
    \begin{array}{c}X\\Y \end{array}
\right)
=
\left(
    \begin{array}{cc}
    a & b \\
    c & d 
    \end{array}
\right) 
\left(
    \begin{array}{c}x\\y\end{array}
\right)
+
\left(
    \begin{array}{c}p\\q\end{array}
\right),
\label{equ:affine}
\end{equation*}
with affine parameters---i.e., $a, b, c, d$ for rotation and deformation correction and $p, q$ for displacement correction.
Here, $p$ and $q$ are defined as relative displacement vectors from one reference sensor to a certain sensor.

\subsection{Sensor alignment using grains recorded in a real emulsion}
\label{sec:calibration_before}
To determine the affine parameters of each sensor, real grains contained in a nuclear emulsion are used as in the calibration of segmented sensors of a telescope using a star catalog.
Calculation of affine parameters is performed as follows.
First, $a_i$, $b_i$, $c_i$, and $d_i$ are obtained by comparing the amount of movement in a sensor coordinate and stage coordinate by moving the stage in the $X$ and $Y$ directions.
Second, a main grain catalog of 5$\times$5~mm$^2$ is created by using a reference sensor,
and subgrain catalogs of 1$\times$0.5~mm$^2$ are created by using the other sensors.
These catalogs are corrected by using $a_i$, $b_i$, $c_i$, and $d_i$ obtained in the above procedure.
Finally, the displacement parameter of $p_i$ and $q_i$ can be obtained by comparing each subcatalog to the main catalog.
These catalogs are revised when the parameters are remeasured.

\begin{figure}[htbp]
\centering
\includegraphics[width=0.5\textwidth]{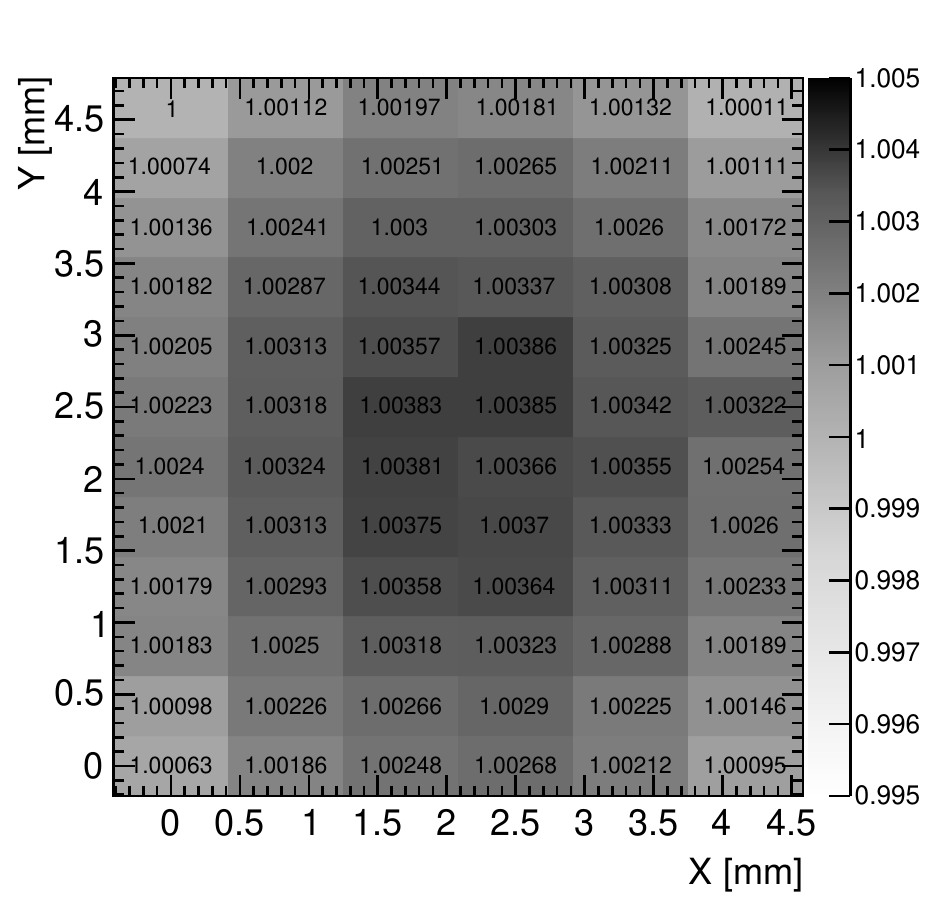}
\caption{Magnification ratio of each sensor. The minimum value appears at the upper-leftmost sensor and is set to one.}\label{fig:sensor_magnification}
\end{figure}

The average magnification ($mag=\frac{1}{2}\left( \sqrt{a^2+b^2}+\sqrt{c^2+d^2}\right)$) is 0.4545~$\rm{\mu m}$/pixel over all sensors.
The ratio of $mag$ to the minimum one is shown in Fig.~\ref{fig:sensor_magnification}.
The maximum $mag$ is 0.39\% larger than the minimum $mag$ by the radial distortion, and these two sensors are 3.0~mm apart.
This corresponds to radial distortion of 0.234\% from the center to the corner of the FOV.
By this value, the distortion of an 80-$\rm{\mu m}$-long track, which is the longest track in the standard setting, is calculated to be 0.1~pixels, which is found to be sufficiently smaller than the required value of one pixel.

\subsection{Calibration method by using real scanning data}
\label{sec:calibration_after}

The affine parameters are sometimes shifted by aging, thermal expansion of the camera support, etc.
A trajectory in the overlap area (27\% in the whole area) between adjacent sensors is then detected at two positions.
We devised and implemented a new method using real tracks in normal data capture because the method described in section~\ref{sec:calibration_before} is quite time consuming.

\begin{figure}[htbp]
\centering
\subfloat[$X$-axis]{\includegraphics[width=0.45\textwidth]{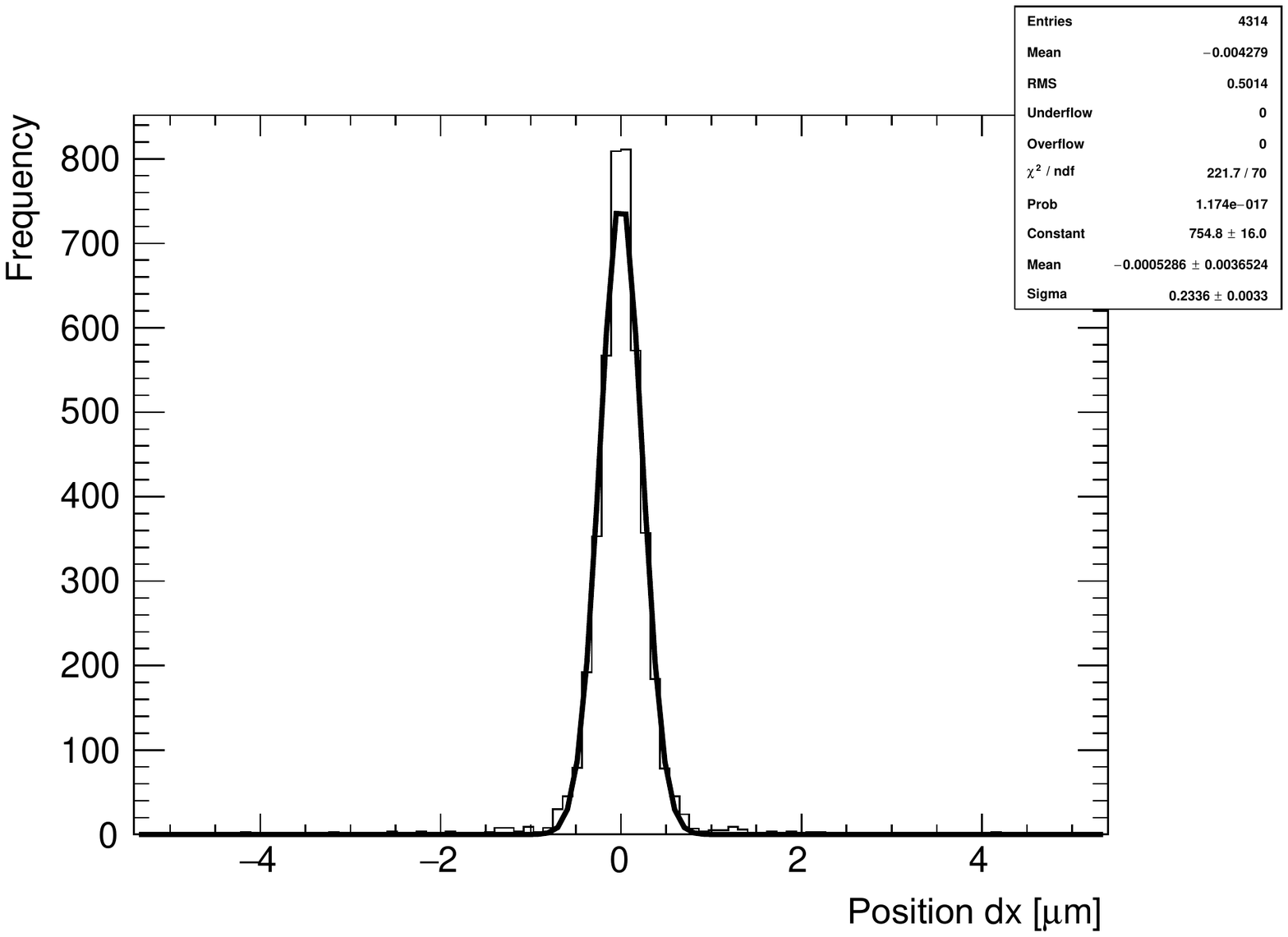}}
\hspace{1em}
\subfloat[$Y$-axis]{\includegraphics[width=0.45\textwidth]{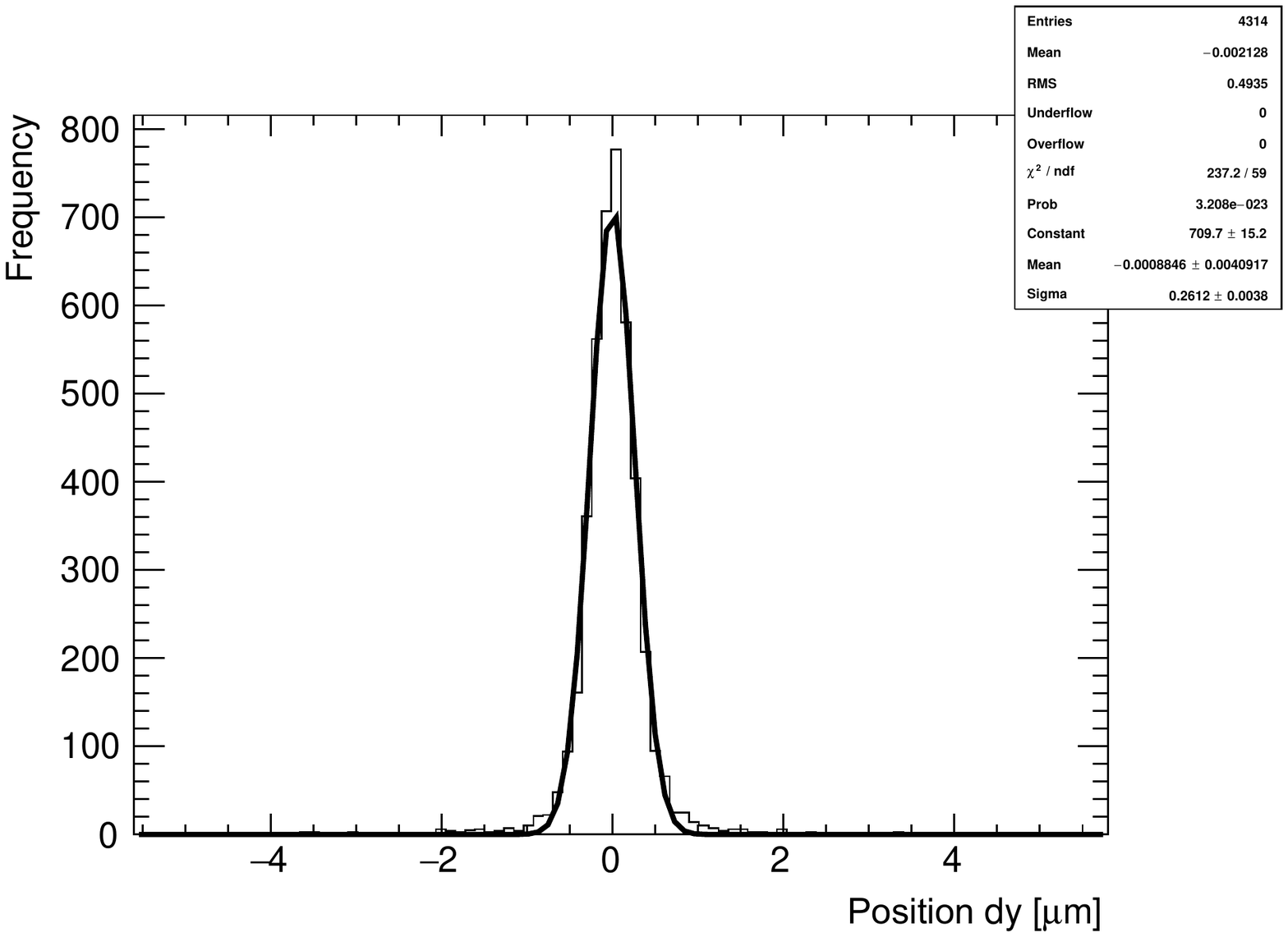}}
\caption{Distribution of position difference of the tracks recognized in the overlap area of the two adjacent sensors.}\label{fig:iti_zure}
\end{figure}

To determine the relative displacement of each sensor with the real tracks, 
the position displacement vector ($D_1$, $D_2$, $D_3$, $D_4$) relative to the four adjacent sensors is calculated for each sensor by comparing the position difference of those simultaneously recognized tracks.
The most probable relative displacement vector $P_{i}(x,y)$ is then calculated by minimizing the residuals $\sigma$ defined as
\begin{equation*}
\sigma^2=\sum_{i=1}^{72}\sum_{j=1}^4 \left( D_{ij} - P_{i} \right)^2,
\end{equation*}
by analytical calculations assuming $P_{1}$ to be $(0, 0)$, where $i$ is the sensor number.
The residuals are calculated in the $X$ and $Y$ directions independently.
After this process, the track positions are converted again with the new affine parameters.
The distribution of the position difference of tracks in all adjacent sensors is then obtained as shown in Fig.~\ref{fig:iti_zure}.
The sigma values of $dx$ and $dy$ are obtained to be 0.23 and 0.26~$\rm{\mu m}$ for the $X$-and $Y$-axes, respectively. 
This accuracy is precise enough for a pixel size of 0.45~$\rm{\mu m}$.

\section{Stage tuning}
Because HTS has utilized step movement (go and stop) to scan films and the distance of the step has been an order of magnitude longer, the vibration effect caused by the stop movement should be well understood.
The residual vibrations after stop may cause blurred grain images or curved tracks, and then the pulse height may be decreased and the recognized position and angle may be shifted.

\subsection{Evaluation of the vibration effect}
To evaluate the vibration, 30 images (=0.1~s) were taken while fixing the $Z$ position from the moment when the encoder of the $X$-axis stage reaches the target position.
Using grains recognized in the captured images, positional deviations between sequential frames were measured.

\subsection{Result}
\begin{figure}[htb]
\centering
\includegraphics[page=2,width=0.6\textwidth]{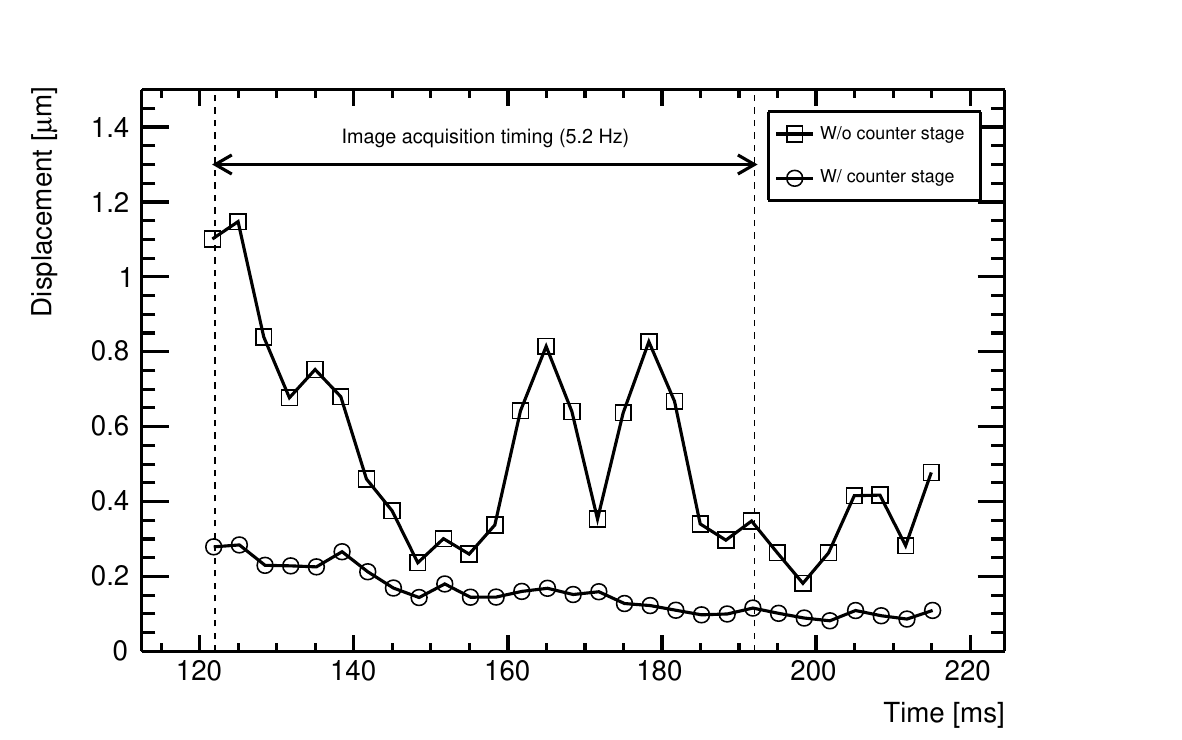}
\caption{
    Effect of the counter stage.
    The circle and square show the vibration in the cases with and without the counter stage movement, respectively.}\label{fig:Stage_acc_disp_wo_counter}
\end{figure}

First, the maximum velocity was fixed at 100~mm/s, and the acceleration was varied from 1.0--10~m/s$^2$.
The counter stage was driven to be synchronized to the main stage.
When the acceleration was greater than 5~m/s$^2$, the vibration was not settled within 0.12~s.
Therefore, we set the acceleration as 5~m/s$^2$.
Next, we evaluated the difference with and without the counter stage movement.
The result is shown in Fig.~\ref{fig:Stage_acc_disp_wo_counter}.
The amount of vibration is significantly reduced with the counter stage movement.

\section{Evaluation}
\subsection{Overview of the performance evaluation}
The performance of an automatic readout machine is evaluated mainly in terms of track finding efficiency and angular accuracy.
We evaluated the HTS performance by using real data of the nuclear emulsion films exposed in a GRAINE 2015 flight~\cite{Ozaki:2015xlw}. The films have a size of 25$\times$38~cm$^2$.
Emulsion layers approximately 60~$\rm{\mu m}$ thick were coated on both sides of a 180-$\rm{\mu m}$-thick polystyrene plastic base.
The density of silver bromide crystals in the emulsion is higher than that of OPERA film~\cite{Nakamura:2006xs}, which was used in S-UTS evaluation.
All films contained in the converter section and the shifter section were readout by the HTS.
For this evaluation, only the bottom three films of the convert section in the area of 117~cm$^2$ were used.
The quality of the converter and shifter section will be discussed in another paper.

In the GRAINE 2015 flight, the nuclear emulsion has a history of 8.5~days on the ground and 14.4~h on the flight.
Level flight at an altitude of 37.2~km was continued for 11.5~h.

\subsubsection{Track reconstruction}
\begin{figure}[htbp]
\centering
\subfloat[]{\includegraphics[page=1,width=0.4\textwidth]{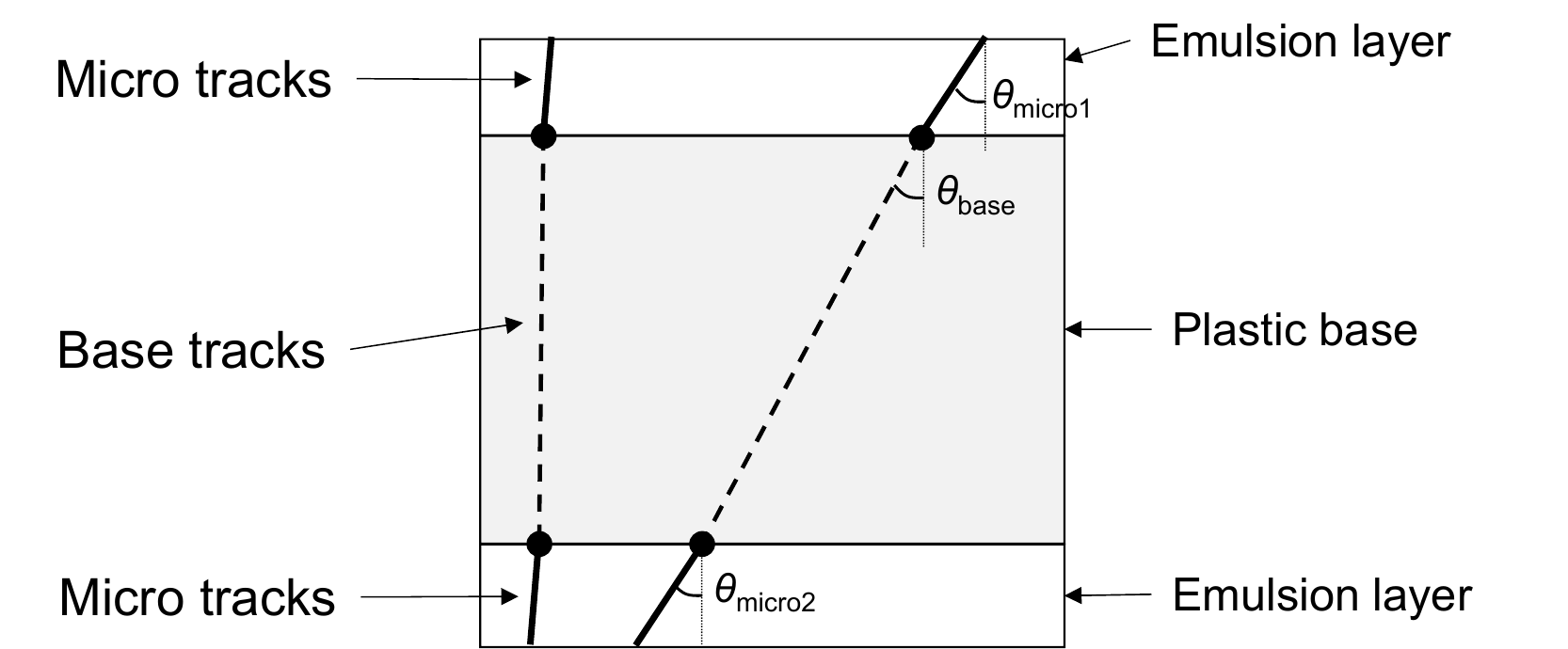}}
\hspace{1em}
\subfloat[]{\includegraphics[page=3,width=0.4\textwidth]{./figs_HTS_figs_base_micro2.pdf}}
\caption{(a) The definition of micro tracks and base tracks. (b) The definition of the angle and position difference between two base tracks. The angle difference is $|\theta_1 - \theta_2|$, and the position difference is $|P_1 - P_2|$.}\label{fig:micro_base_tracks}
\end{figure}

Micro tracks of two emulsion layers on both sides of a plastic base were connected, and then a base track candidate was selected.
As shown in~\ref{fig:micro_base_tracks} (a), we connected the edges of two micro tracks on the base side to establish the angle and position of the base track.
The PH of the base track is the summed PH of two micro tracks.
We have defined that a more likely base track has a smaller angular difference between each micro track and the base track.

Because the angle of a micro track $\overrightarrow{ma}$ is affected by the shrinkage and the distortion of the emulsion layer caused by the development, 
those effects should be corrected by using the so-called shrinkage factor $shr$ and distortion vector $\overrightarrow{dist}$ as follows:
\begin{eqnarray*}
\begin{split}
\label{eq:dc_correction}
\overrightarrow{ma}' = shr \cdot \overrightarrow{ma}  + \overrightarrow{dist}.
\end{split}
\end{eqnarray*}
where $\overrightarrow{ma}$ is the original angle, and $\overrightarrow{ma}'$ is the corrected angle.
The shrinkage factor and distortion vector are calculated to maximize the number of connected base tracks. 
The base tracks crossing multiple films are reconstructed with the angle and position consistency as shown in Fig.~\ref{fig:micro_base_tracks} (b).

\subsection{Results}
\subsubsection{Base track finding efficiency}
To evaluate the track finding efficiency at a film, tracks were reconstructed by using the films sandwiching the target film, and the existence of the reconstructed tracks was checked at the target film.
The base track finding efficiency of a film is defined as the ratio between the number of found predicted tracks and the number of all predicted tracks.

\begin{figure}[htbp]
\centering
\includegraphics[width=0.6\linewidth]{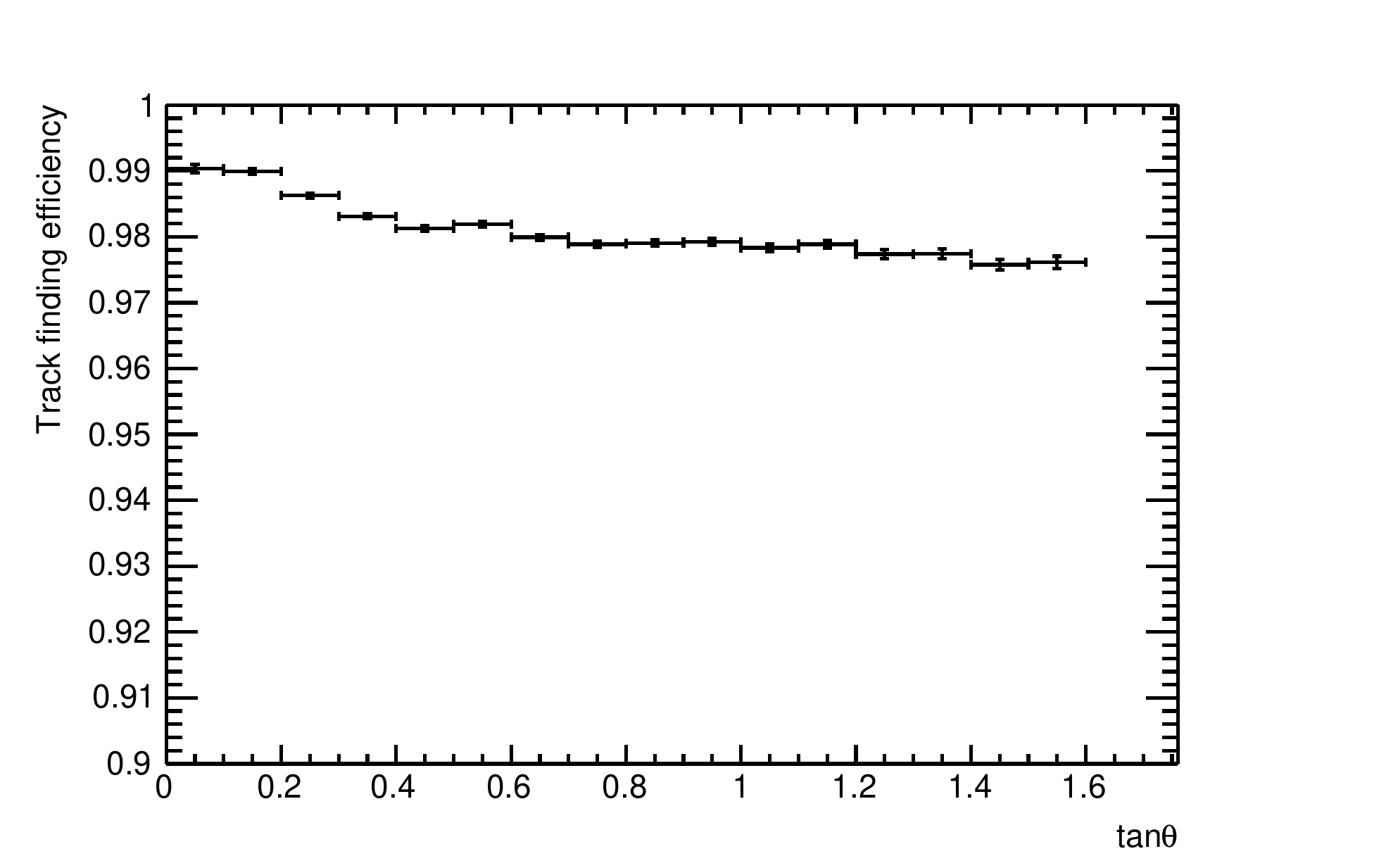}
\caption{Angle dependence of the track finding efficiency of base tracks.}
\label{fig:efficiency2}
\end{figure}

\begin{figure}[htbp]
\centering
\subfloat[]{\includegraphics[page=2,width=0.45\textwidth]{./figs_ph_mt_bt_summed_ph_mt.pdf}}
\hspace{1em}
\subfloat[]{\includegraphics[page=1,width=0.45\textwidth]{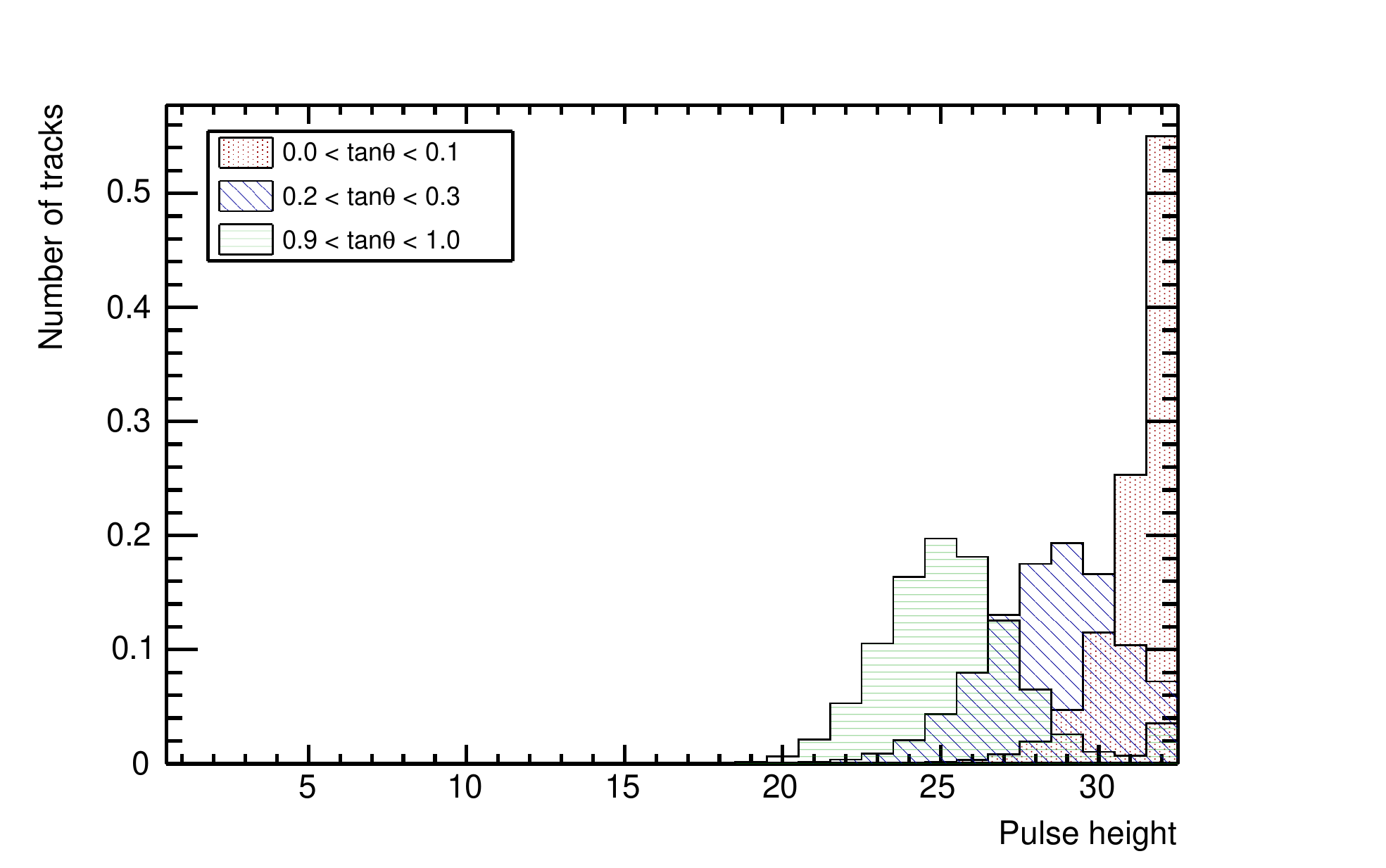}}
\caption{(a) Angle dependence of average pulse height of the base tracks. The error of the average is a standard deviation. (b) Pulse height distributions for the base tracks of $0.0\!<\!\tan \theta\!<\!0.1$, $0.2\!<\!\tan \theta\!<\!0.3$, and $0.9\!<\!\tan \theta\!<\!1.0$.}\label{fig:ph_angle}
\end{figure}

The measured angular dependence of the track finding efficiency is shown in Fig.~\ref{fig:efficiency2}.
The efficiency was greater than 97\% in the angular range of $\tan \theta\!<\!1.6$.
The efficiency for $\tan \theta$ less than 0.2 is slightly higher than for the other regions.
The angle dependence of the pulse height is shown in Fig.~\ref{fig:ph_angle} (a).
The efficiency is correlated to the average PH.
The PH distributions of the found tracks for $0.0\!<\!\tan \theta\!<\!0.1$, $0.2\!<\!\tan \theta\!<\!0.3$, and $0.9\!<\!\tan \theta\!<\!1.0$ are shown in Fig.~\ref{fig:ph_angle} (b).
The PH averages for $0.0\!<\!\tan \theta\!<\!0.1$, $0.2\!<\!\tan \theta\!<\!0.3$ and $0.9\!<\!\tan \theta\!<\!1.0$ were measured to be 31.2, 28.5, and 25.4, respectively.

\subsubsection{The angular accuracy}
\begin{figure}[htbp]
\centering
\includegraphics[page=1,width=1.0\linewidth]{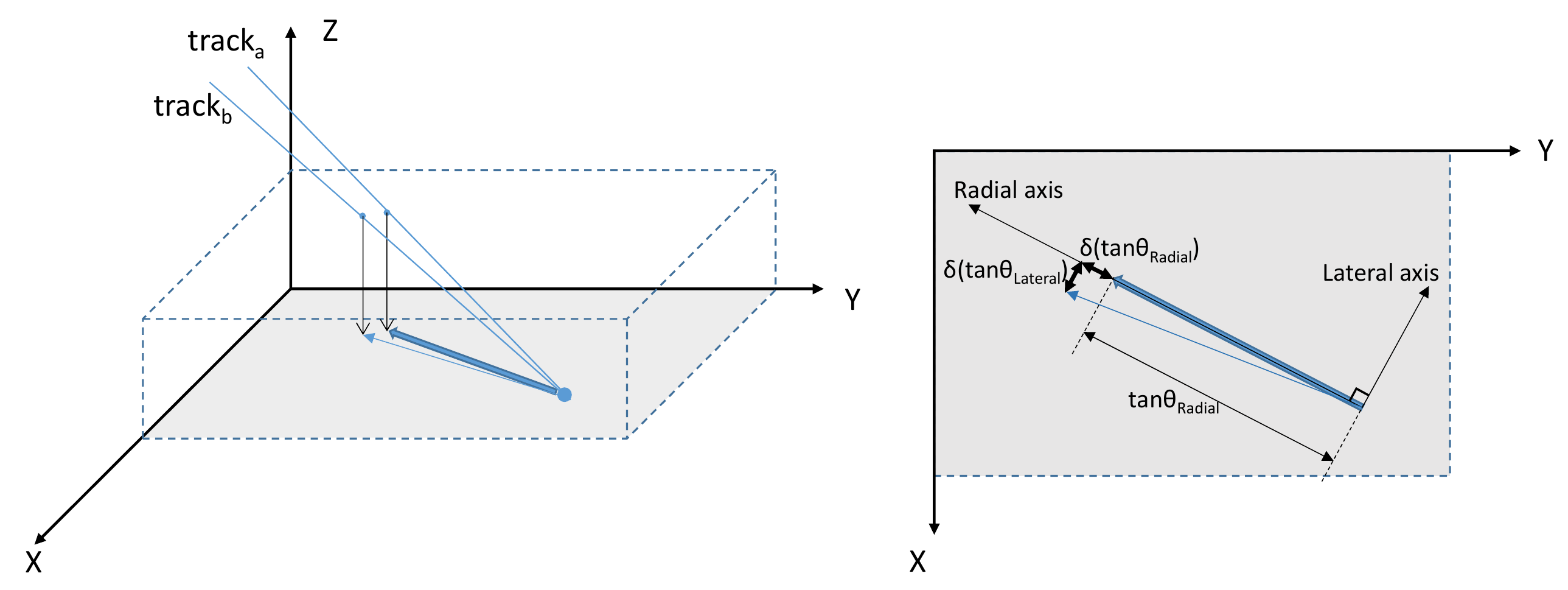}
\caption{Schematic view of the ``Radial-Lateral'' component definition of angular difference between two tracks.}
\label{fig:radial_lateral}
\end{figure}

\begin{figure}[htbp]
\centering
\includegraphics[page=2,width=0.6\linewidth]{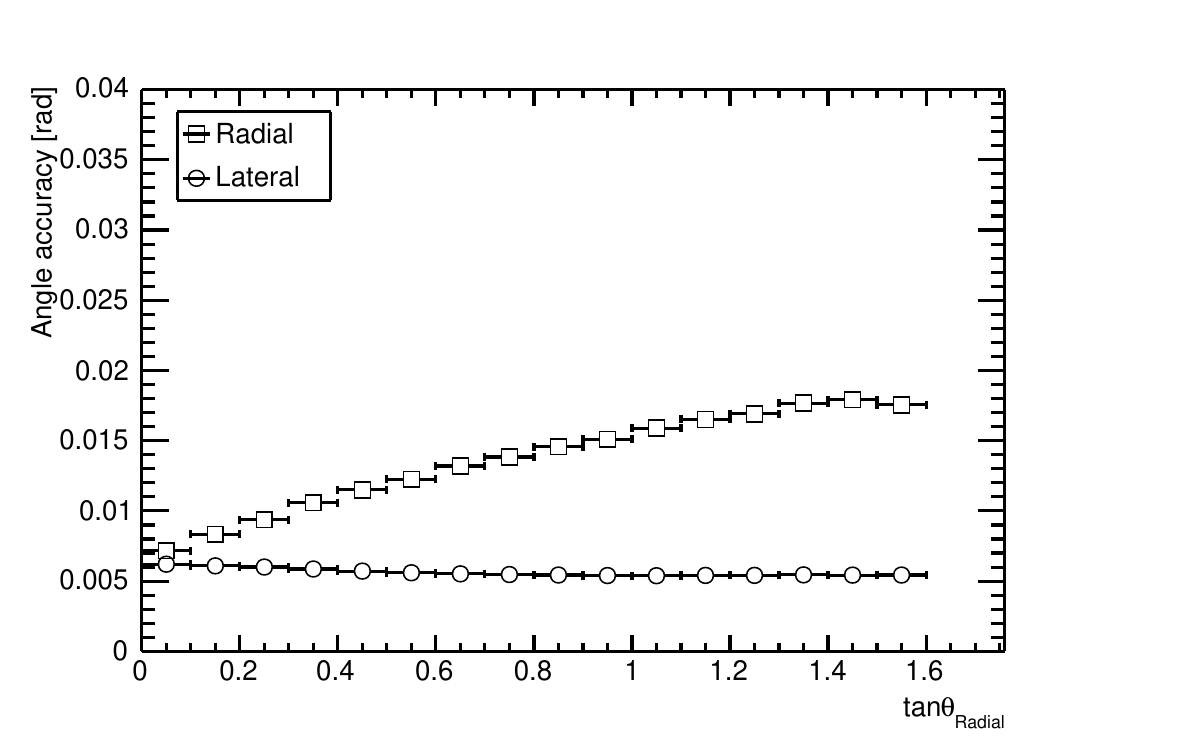}
\caption{Angle dependence of angle accuracy for micro tracks.}
\label{fig:angle_micro}
\end{figure}

\begin{figure}[htbp]
\centering
\includegraphics[width=0.6\linewidth]{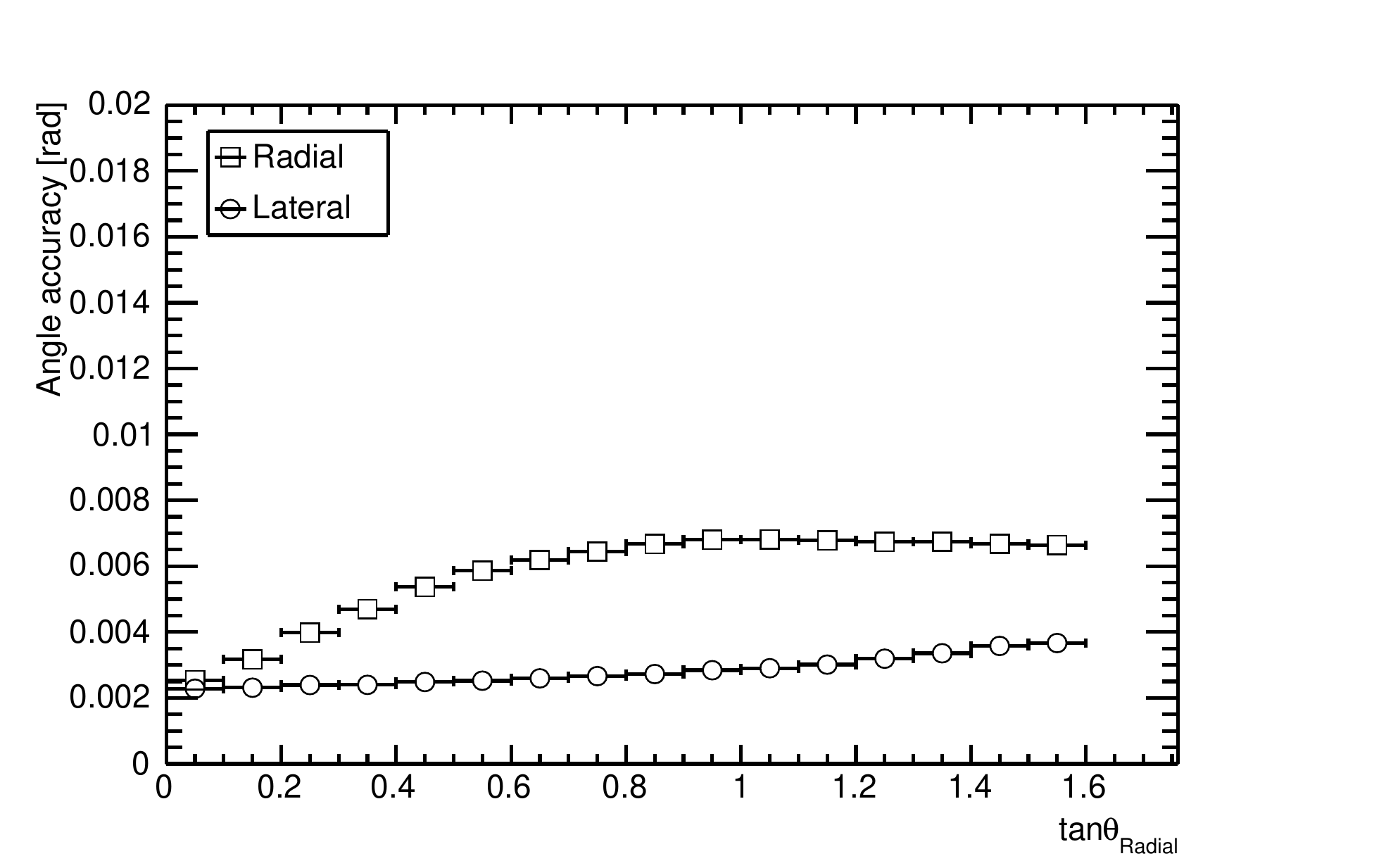}
\caption{Angle dependence of angle accuracy for base tracks.}
\label{fig:angle_base}
\end{figure}

The angular accuracy of the micro tracks is defined as the angular difference between the micro tracks and the base tracks, 
and that of base tracks is defined as the angular difference between the base track in the evaluating film and that in the next film.
As shown in Fig.~\ref{fig:radial_lateral}, the angular accuracy was resolved for ``radial and lateral'' components;
the radial axis is the horizontal direction of the track, and the lateral axis is the perpendicular direction.
Because the depth of field is greater than the lateral resolution, the two components are very different in the large angle tracks.
The angle dependence for the angular accuracy of the micro tracks is shown in Fig.~\ref{fig:angle_micro}, and that of the base tracks is shown in Fig.~\ref{fig:angle_base}.

\subsubsection{The track finding efficiency dependence of readout repetition frequency}

The repetition frequency in the above evaluation was 4.2~view/s, which corresponds to a readout speed of 3800~cm$^2$/h.

\begin{figure}[htbp]
\centering
\includegraphics[ width=0.6\textwidth]{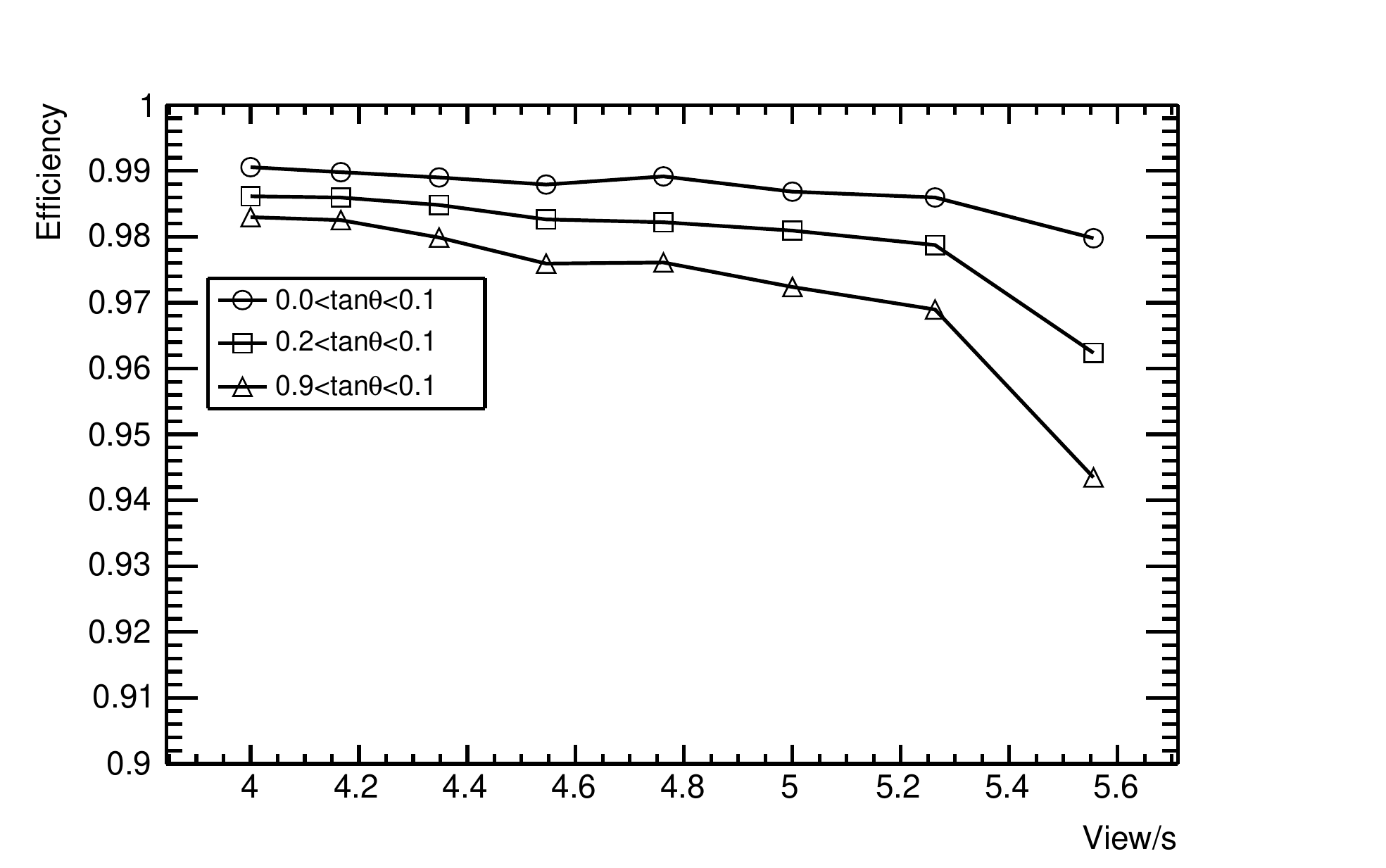}
\caption{
    Dependence of the track finding efficiency for the readout speed.
    Here, 5.2~view/s corresponds to 4700~cm$^2$/h.}
\label{fig:efficiency_speed}
\end{figure}

We investigated the track finding efficiency as a function of the repetition frequency from 4~view/s to 5.5~view/s.
The frequency was changed by changing the timing of the image data capture, and the peak speed and acceleration of the $X$-axis stage were not changed.
The result is shown in Fig.~\ref{fig:efficiency_speed}. 
The efficiency becomes lower when the frequency becomes higher.
For example, the efficiency was deteriorated by 1\% in the case of the track with an angle of $0.9\!<\!\tan \theta\!<\!1.0$ when the frequency was changed from 4.0~view/s to 5.2~view/s.
If a 1\% decrease is acceptable, a readout speed of 4700~cm$^2$/h at 5.2~view/s can be executed.

\subsection{Discussion}

The pulse height of the smaller angle track is higher because of the focal depth.
The distance in the $Z$ direction between two adjacent captured images is determined to be almost the same as the focal depth of the HTS, approximately 4~$\rm{\mu m}$.
When two grains are lined up in the $Z$ direction---i.e., optical axis---the grains spread over images of the different depths.
As a result, it is considered that the PH is amplified at a smaller angle.

Angular accuracies of micro tracks in the range of $|\tan \theta|\!<\!0.1$ are 7~mrad in HTS and 13~mrad in S-UTS~\cite{Morishima:2010zz}, and 
those in the range of $0.4\!<\!|\tan \theta|\!<\!0.5$ are 11~mrad in HTS and 23~mrad in S-UTS. 
Although it is difficult to compare directly owing to the different thicknesses of the emulsion layer, the angular accuracy of HTS seems better than that of S-UTS. 

The angular accuracies of base tracks in the range of $|\tan \theta|\!<\!0.1$ are 2.5~mrad in HTS and 2.5~mrad in S-UTS, and those in the range of $0.4\!<\!|\tan \theta|\!<\!0.5$ are 5.4~mrad in HTS and 4.2~mrad in S-UTS. 
Although it is also difficult to compare directly owing to the difference in sensitivity to charged particles, incident particles, thickness of plastic base, uniformity of thickness, and variation in shrinkage and distortion, 
the angular accuracy of HTS in the small angle is almost the same as that of S-UTS, and that of HTS in the larger angle seems slightly lower.

To improve the angular accuracy of the base track even at a large angle, it is necessary to improve the position accuracy in the $Z$ direction.
When acquiring images, because the clocks of 72 sensors are not synchronized, the focal plane has an error of approximately 4~$\rm{\mu m}$ for capturing images.
Furthermore, the image of one grain has an error of 4~$\rm{\mu m}$ in the depth direction because of the focal depth.
Implementation of synchronous signals and advances in image processing will solve the angular accuracy problem.

At the HTS, we have improved the readout speed by widening the FOV.
As described in section~\ref{sec:Concept}, to increase the readout speed, there are two directions; one is to increase the frequency of readout repetition, and the other is to widen the FOV.
HTS took the second way.
For further speedup, we are designing a new readout system, HTS-2, along both directions.
An objective lens with a double-wider FOV and contentious movement with a diagonal focal plane will be adapted.
HTS-2 will hopefully achieve a scanning speed of 25,000~cm$^2$/h, which is approximately 5 times faster than that of the HTS (HTS-1).

\section{Conclusion}
Nuclear emulsion is a three-dimensional particle tracking detector with more than 100 years of history, but it still has an extensive application field thanks to the invention of automatic nuclear emulsion readout systems.
The newest machine, HTS, described in this paper is intended to address the modern applications.
HTS has been developed with a wide field lens of 5.1$\times$5.1~mm$^2$ FOV and 72 two-megapixel sensors to cover this FOV.
We also applied 72 GPUs and 36 CPUs for image processing.
Scanning at 5.2~view/s with 5~mm step movement has been achieved by utilizing the counter stage.
Finally, HTS achieved a readout speed of approximately 0.5~m$^2$/h, which is almost two orders faster than that of the previous system used in the OPERA experiment.
This speed corresponds to a scanning area of $\sim$1000~m$^2$/year.

There are two main subjects for HTS.
One is a wide field parallel readout by multiple sensors maintaining submicron accuracy.
The other is vibration suppression after the step movement.
As described above, the new method for multisensor position alignment has been accomplished by using real grain and track data.
For vibration suppression, the realistic acceleration value was tuned by using a counter stage.

The previous HTS readout area of approximately 50~m$^2$ in the GRAINE 2015 flight showed that the tracking efficiency is greater than 97\% for the tracks with an angle of $\tan\theta\!<\!1.0$ and contributed to the observation of cosmic gamma ray.
Relating to muon radiography, an area of approximately 80~m$^2$ was readout, and new inner structures were identified in pyramids.
In the NINJA project, HTS readout an area of approximately 20~m$^2$ and contributed to the study of low-energy neutrino interactions.
In addition, the availability of HTS stimulates new projects and triggers new collaborative work, and HTS is becoming an indispensable tool for future radiation measurements. 

\section*{Acknowledgment}
We would like to thank T.~Kawai for constructing HTS.
We wish to acknowledge valuable discussion on the calibration method with S.~Kukita.
This work was supported by JSPS KAKENHI Grant Numbers JP22340057 and the JST-SENTAN Program from Japan Science and Technology Agency, JST.
Some instruments were supplied by Kobayashi-Maskawa Institute for the Origin of Particles and the Universe (KMI).
M.~Yoshimoto and H.~Kawahara were supported by a Grant-in-Aid for JSPS Research Fellow.
I thank all collaborators who have utilized HTS for their experiments.

% can use a bibliography generated by BibTeX as a .bbl file
% BibTeX documentation can be easily obtained at:
% http://www.ctan.org/tex-archive/biblio/bibtex/contrib/doc/

% bibtex Author_tex
% \bibliographystyle{ptephy}
% \bibliography{ref}
%
% once the .bbl file has been generated then place the text in your article.

\end{document}